\documentclass[aps,pra,twocolumn,showpacs,superscriptaddress]{revtex4-1}

\usepackage{graphicx}
\usepackage{amsmath}  
\usepackage{amsthm}  
\usepackage{amssymb}  
\usepackage{color}
\usepackage{hyperref}
\usepackage{longtable}
\usepackage{multirow}
\usepackage{mathrsfs}
\usepackage{braket}

\let\savedegree\corresponds
\let\corresponds\relax
\usepackage{mathabx}
\let\corresponds\savedegree

\usepackage{float}
\usepackage{bm}
\usepackage[percent]{overpic}
\usepackage{mathrsfs}
\usepackage{mathcomp}
\usepackage[table,dvipsnames]{xcolor}

\definecolor{red}{rgb}{1.0, 0.7, 0.7}
\definecolor{blue}{rgb}{0.7, 0.7, 1.0}

\begin{document}
\title{Multipartite-entanglement detection with projective measurements}
	
\author{Arun Sehrawat}
\email[email: ]{aruns@iisermohali.ac.in}
\affiliation{Department of Physical Sciences, 
Indian Institute of Science Education \& Research (IISER) Mohali, 
 Sector 81 SAS Nagar, Manauli PO 140306, 
Punjab, India}


\begin{abstract}
For a projective measurement, the Born rule provides the probability
for an outcome in terms of the inner product between a projector and a quantum state. If the projector represents a pure entangled state and the state for a composite system is separable, then we cannot get probability 1 for the outcome. This insight delivers a single condition for entanglement detection. By applying local unitary transformations from the Clifford group, we turn one condition into many. Furthermore, we present two equivalent schemes---one employs global and other requires local projective measurements---to test these conditions in an experiment.
Here, a global measurement is characterized by an orthonormal basis that holds local-unitary-equivalent entangled kets. 
Whereas a local-measurement setting is specified by
mutually unbiased bases assigned to the subsystems.
We also supply a straightforward (computer) algorithm to generate all the conditions and then to check whether a state is shown 
entangled or not by these conditions.
Finally, we demonstrate every element of our schemes by considering several well-known examples of entangled kets and states.
\end{abstract}


\maketitle

\section{Introduction}\label{sec:Intro}

Quantum entanglement \cite{Horodecki09} can provide stronger correlations than all those belong to the classical realm, and it is due to the entanglement certain information-processing tasks \cite{Ekert91,Bennett93} come to a realization. 
Therefore, creation, detection, qualification, and protection of this
newly discovered resource become crucial. 
Here we are focusing entirely on the detection, for which
several methods based on the positive partial transpose \cite{Peres96,Horodecki96},
entanglement witnesses \cite{Horodecki96,Terhal00,Lewenstein00,Acin01,Bruss02,Wei03,
Bourennane04,Toth05,Guhne05,Toth05b,Toth07,Yu05,Terhal01}, uncertainty measures \cite{Horodecki96b,Hofmann03,Guhne04,Guhne04b,Giovannetti04}, and
correlation functions \cite{Kothe07,Maccone15}
are proposed (for a review, see \cite{Guhne09}).

Generally, rules for the detection are laid out in terms of certain mathematical inequalities that carry expectation values of some operators. Violation of (or, in some cases, satisfying) such an inequality reveals entanglement among the constituents of a composite system.
We present two such schemes for multipartite-entanglement detection
in Sec.~\ref{sec:criteria} accompanied with a few renowned \cite{Einstein35,Bohm51,Werner89,Horodecki97,Horodecki98,Horodecki01,Bennett99,DiVincenzo03,Terhal01,Greenberger89,Greenberger90,Hein06,Due00} and recent \cite{Rossi13,Guhne14} examples of entangled kets and states in Sec.~\ref{sec:Examples}.
Since local (attached to an individual subsystem) unitary operations do not change the amount and nature of entanglement,
all the kets that are locally equivalent---which means obtained by applying local unitary transformations alone---to an entangled ket belong to the same class.

In various cases---like the examples in Sec.~\ref{sec:Examples}---a set of locally-equivalent entangled kets constitutes an orthonormal basis of Hilbert space affiliated with the composite system.
Such a basis, called \emph{entangled basis} (for instance, see in \cite{Bennett93,Horodecki96b,Guhne04,Guhne04b,Giovannetti04,Lawrence02,Klimov07}), defines a global projective measurement on the system. 
Whether or not an entangled basis exists for an entangled ket, one can fully utilize our techniques for the detection.
Our methods are founded on the following three things.

First, if a compounded system is in a separable state, 
then one cannot get every time the same outcome (that is probability one) when the system is measured in an entangled basis.  
This fact brings us inequalities, one for each projector onto a ket of the basis, violation of which leads to the detection.
Such an entangled projector is related to a witness operator based on the geometric measure of entanglement \cite{Wei03} (see also witness operators in \cite{Bourennane04,Acin01,Toth05,Guhne04b,Guhne05,Toth05b,Toth07,Terhal01}), and the global measurement in an entangled basis realizes a set of (mutually exclusive) witness operators.

Second, the generalized Pauli---also known as Heisenberg-Weyl---operators constitute an orthogonal basis, called \emph{Pauli basis}, of the Hilbert-Schmidt operator space \cite{Weyl32,Schwinger60,Englert06} for every subsystem.
Then, tensor products of the Pauli operators assemble \emph{product-Pauli basis} for the composite system. 
Hence every entangled projector---even if it does not come from an entangled basis---can be decomposed into a linear combination of product-Pauli operators that can be viewed as \emph{elementary-correlation operators}.
This translates our detection conditions in terms of correlation functions, which can be estimated by certain local projective measurements rather than the global.
Now one can see the existence of an entangled basis is not an inevitable
requirement for us.  
Our single local-measurement setting is simply a collection of
mutually unbiased bases (MUBs) \cite{Ivanovic81,Durt10}, one basis for each subsystem.

Third, a unitary operator of the Clifford group transforms one Pauli operator into another under unitary conjugation \cite{Gottesman98,Gottesman99}, named \emph{Clifford conjugation}.
So one can obtain many more conditions with these conjugations from a single condition provided in terms of the product-Pauli operators.
Every conjugation is achieved here with a tensor product of local Clifford operators.
To gain more conditions, we are not using non-product (that cannot be broken into a tensor product) Clifford operators as they can change entanglement.
In Appendix~\ref{app:Pauli Op}, we compile all the necessary details from Refs.~\cite{Lidl86,Weyl32,Schwinger60,Ivanovic81,Englert01,Bandyopadhyay02,Lawrence02,Klimov05,Klimov07,Durt10,Englert06,Gottesman98,Gottesman99,Wootters90} for this article about the Pauli group, the Clifford group, and MUBs.
We conclude the article with Sec.~\ref{sec:Conc}.


\section{Entanglement-detection schemes}\label{sec:criteria}

Suppose our quantum system is made of $N$ subsystems, where $i$th subsystem is of $d_i$ levels and 
every $d_i$ is (not necessarily distinct but) a \emph{prime} number.
As there is no further partition of the joint system, one can not do better in terms of entanglement detection by merely changing the partition. 
If $\mathscr{H}_{d_i}$ is the Hilbert space for $i$th constituent, then ${\mathscr{H}_\mathsf{d}=\operatorname*{\otimes}_{i=1}^{N}\mathscr{H}_{d_i}}$ of dimension ${\mathsf{d}=\textstyle\prod\nolimits_{i=1}^{N}d_i}$ is for the compounded system.
In this section, we present two equivalent schemes---one employs global and other adopts local projective measurements---to detect entanglement among the $N$ components.
Throughout the paper, word `local' signifies that a mathematical object is associated with an individual subsystem and word `product' indicates the tensor product of objects of the same kind. 
For example, $L_i$ is a local operator that acts on kets ${|\psi_i\rangle\in\mathscr{H}_{d_i}}$, then
any ket and operator of the forms
\begin{equation}
\label{pro}
|\Psi\rangle:=
\operatorname*{\otimes}_{i=1}^{N}|\psi_i\rangle
\quad\mbox{and}\quad
\mathsf{L}:=
\operatorname*{\otimes}_{i=1}^{N}L_i
\end{equation}
are called \emph{product ket} and \emph{product operator}, respectively.
Furthermore, an orthonormal basis of $\mathscr{H}_\mathsf{d}$ is called \emph{product basis} if and only if its every member is a product ket [for example, see \eqref{pro-basis}].

If ${|\textsc{e}\rangle\in\mathscr{H}_\mathsf{d}}$ is not a product ket then it---is an entangled ket---belongs to
an entanglement class.
Having a set of product-unitary operators 
\begin{equation}
\label{local}
\mathcal{L}_\textsc{e}:=
\big\{\mathsf{L}_\mathsf{k}\big\}_{\mathsf{k}=
	\mathsf{0}}^{\mathsf{d}-\mathsf{1}}
\quad\mbox{such that}\quad
\langle\textsc{e}|\mathsf{L}^\dagger_{\mathsf{k}'}\mathsf{L}_\mathsf{k}
|\textsc{e}\rangle
=\delta_{\mathsf{k},\mathsf{k}'}
\end{equation}
for every $\mathsf{k}$ and $\mathsf{k}'$, we acquire an entangled basis
\begin{equation}
\label{Ent-basis}
\mathfrak{B}_\textsc{e}:=
\big\{|\textsc{e}_\mathsf{k}\rangle\big\}%
_{\mathsf{k}=\mathsf{0}}^{\mathsf{d}-\mathsf{1}}\,,
\quad \mbox{where}\quad
|\textsc{e}_\mathsf{k}\rangle:=
\mathsf{L}_\mathsf{k}\,|\textsc{e}\rangle\,
\end{equation}
(for example, see in \cite{Bennett93,Horodecki96b,Guhne04,Guhne04b,Giovannetti04,Lawrence02,Klimov07}).
It is an orthonormal basis of Hilbert space $\mathscr{H}_\mathsf{d}$ due to the relation~\eqref{local}, where $\delta_{\mathsf{k},\mathsf{k}'}$ is the Kronecker delta function.
As quantum entanglement remains intact in quality and quantity after application of any product-unitary operator, each ket of $\mathfrak{B}_\textsc{e}$---is locally equivalent to another---represents the same entanglement as ${|\textsc{e}\rangle}$ does.
Without loss of generality, we take $\mathsf{L}_\mathsf{0}$ as 
the identity operator on $\mathscr{H}_\mathsf{d}$, and then ${|\textsc{e}_\mathsf{0}\rangle=|\textsc{e}\rangle}$.

Set of all (bounded) operators ${\mathscr{B}(\mathscr{H}_\mathsf{d})}$ that are defined on $\mathscr{H}_\mathsf{d}$ constitutes a $\mathsf{d}^2$-dimensional Hilbert-Schmidt space with
the inner product
\begin{equation}
\label{HS-inner-pro}
\lgroup A,B\,\rgroup_{\textsc{hs}}=
\text{tr}(A^\dagger B)\,,
\quad\mbox{where}\quad A,B\in\mathscr{B}(\mathscr{H}_\mathsf{d})\,.
\end{equation}
In quantum theory, state for a $\mathsf{d}$-level system is represented
by a positive operator ${\rho=\rho^\dagger\geq0}$ with ${\text{tr}(\rho)=1}$,
and the expectation value of an operator $O$ is given by
\begin{equation}
\label{Exp-O}
\langle O\rangle_{\rho}=
\text{tr}(\rho\, O)=\lgroup\rho,O\rgroup_{\textsc{hs}}\,.
\end{equation}
Operators that we pick here are orthogonal-projection operators (projectors), ${\varPi=\varPi^\dagger=\varPi^2}$, of rank 1, which means ${\text{tr}(\varPi)=1}$.
At the end of Sec.~\ref{subsec:N-qubit}, in Remark~10, the entangled projector $\varPi_\textsc{upb}^\perp$ is of rank 4 and is 
not onto a single entangled ket, but onto a subspace.

As $\mathfrak{B}_\textsc{e}$ is not a product basis, like $\mathcal{B}_0$ of \eqref{pro-basis}, it specifies a global projective measurement with the projectors
\begin{equation}
\label{prob}
\varPi_\mathsf{k}:=
|\textsc{e}_\mathsf{k}\rangle\langle\textsc{e}_\mathsf{k}|\,,
\quad\text{where}\quad
p_\mathsf{k}=\langle \varPi_\mathsf{k}\rangle_\rho
\quad\text{(Born rule)}\quad
\end{equation}
is the probability of getting $\mathsf{k}$th outcome.
These probabilities constitute a vector ${\vec{p}:=(p_\mathsf{0},\cdots,p_{\mathsf{d-1}})}$, 
which belongs to the probability space $\Omega$ that is defined by
\begin{eqnarray}
\label{p-const1}
\textstyle\sum\nolimits_{\mathsf{k=0}}^{\mathsf{d-1}}
p_\mathsf{k}&=&1\quad\ \mbox{and} \\  
\label{p-const2}
0&\leq&p_\mathsf{k}\quad \mbox{for all} \quad 
\mathsf{0}\leq \mathsf{k}\leq \mathsf{d-1}\,.
\end{eqnarray}
These two statements simply announce that all the probabilities sum up to one and are nonnegative numbers.
The space $\Omega$ is---the standard ${(\mathsf{d-1})}$-simplex---a compact convex subset of the $\mathsf{d}$-dimensional Euclidean space $\mathbb{R}^\mathsf{d}$.
Every point of such a subset can be written as a convex combination of its extreme points due to the Krein-Milman theorem (see Theorem~3.3.5 and Appendix~A.3 in \cite{Niculescu93}).

Only the pure state ${\rho=|\textsc{e}_\mathsf{k}\rangle\langle\textsc{e}_\mathsf{k}|}$  
gives ${p_\mathsf{k}=1}$, which specifies an extreme point of $\Omega$.
There are $\mathsf{d}$ such points, one for each ket in the
basis~\eqref{Ent-basis}. 
Since $|\textsc{e}_\mathsf{k}\rangle$ is not a product ket, like
${|\Psi\rangle}$ in \eqref{pro}, ${\rho=|\Psi\rangle\langle\Psi|}$ \emph{cannot} deliver ${p_\mathsf{k}=1}$ by the Born rule~\eqref{prob}.
So, for every pure product state, we have
\begin{eqnarray}
\label{p<P_E}
p_\mathsf{k}\leq P_\textsc{e}\quad \mbox{for all} \quad 
\mathsf{0}\leq \mathsf{k}\leq \mathsf{d-1}\,,
\end{eqnarray}
where
\begin{equation}
\label{P_E}
P_\textsc{e}:=
\operatorname*{max}_{|\Psi\rangle\,\in\,\mathscr{H}_\mathsf{d}}
|\langle\textsc{e}|\Psi\rangle|^2
\end{equation}
is the maximum overlap between the entangled ket and a product ket.
In \cite{Wei03,Bourennane04}, $P_\textsc{e}$ is computed
for several well-known entangled kets.
Importantly, $P_\textsc{e}$ always lies in the interval ${\big[\tfrac{1}{\mathsf{d}},1\big)}$.

By the definition, every separable (as well as mixed product) state is a convex combination of pure product states \cite{Werner89}:
\begin{equation}
\label{sep-state}
\rho_{\text{sep}}=\textstyle\sum\nolimits_{l}w_l|\Psi_l\rangle\langle\Psi_l|
\quad\qquad (0\leq w_l\,,\
\textstyle\sum\nolimits_{l}w_l=1)\,,
\end{equation}
where each $|\Psi_l\rangle$ is of the form~\eqref{pro}.
For $\rho_{\text{sep}}$, we get the convex combination 
${\vec{p}_{\text{sep}}=\textstyle\sum\nolimits_{l}w_l
	\vec{p}_l}$ of probability-vectors $\vec{p}_l$ that are
associated with the states $|\Psi_l\rangle\langle\Psi_l|$ by the Born rule~\eqref{prob}.
Since each component of every $\vec{p}_l$ follows the restriction~\eqref{p<P_E}, each component of $\vec{p}_{\text{sep}}$ will obey the same.  
Now we present our entanglement-detection test:
\begin{equation}
\label{criterion}
\parbox{0.8\columnwidth}
{
	if $\vec{p}$ associated with a quantum state $\rho$ violates 
	one of the constraints~\eqref{p<P_E}, 
	then $\rho$ is an entangled state.  
}
\end{equation} 
This criterion for detection is sufficient, but not necessary, because certain entangled states follow all the conditions in~\eqref{p<P_E}.
In other words, a single measurement setting $\mathfrak{B}_\textsc{e}$ is
not enough to detect all entangled states [for example, see Sec.~\ref{sec:Examples}].

Let us call collection of all the probability-vectors ${\vec{p}\in\Omega}$ that comply with~\eqref{p<P_E} \emph{separable set} $\mathcal{S}_\textsc{e}$.
It is---a proper subset of the probability space $\Omega$---bounded by $2\mathsf{d}$ 
hyperplanes that are characterized by the \emph{equalities} (${p_\mathsf{k}=0}$ and ${p_\mathsf{k}=P_\textsc{e}}$) in constraints~\eqref{p-const2} and \eqref{p<P_E}.
Furthermore, $\mathcal{S}_\textsc{e}$ is also a compact convex subset of $\mathbb{R}^\mathsf{d}$.
Suppose $m$ is the greatest integer between 1 and $\mathsf{d}$ such that
${mP_{\textsc{e}}\leq1}$, that is ${(m+1)P_{\textsc{e}}>1}$, then
\begin{eqnarray}
\label{ext-pt}
\vec{e}_{m}&=&
\big(\overbrace{P_{\textsc{e}}\,,\cdots,P_{\textsc{e}}}^{m\ \text{times}}\,,\,1-mP_{\textsc{e}}\,,
0\,,\cdots,0\big)
\end{eqnarray}
represents an extreme point of $\mathcal{S}_\textsc{e}$.
One can check that $\vec{e}_{m}$ respects every limitation~\eqref{p-const1}--\eqref{p<P_E}, in particular, it obeys 
$m$ and ${\mathsf{d}-m-1}$ (${\mathsf{d}-m}$ if ${mP_{\textsc{e}}=1}$) number of equality constraints of
type \eqref{p-const2} and \eqref{p<P_E}, respectively.
Since conditions~\eqref{p-const1}--\eqref{p<P_E} are same for every
$\mathsf{k}$, permutations of the coordinates of $\vec{e}_{m}$ deliver all other extreme points.
These are
\begin{equation}
\label{no.ex-pts}
\parbox{0.85\columnwidth}
{
	$\tfrac{\mathsf{d}!}{m!(\mathsf{d}-m-1)!}$ 
	if ${mP_{\textsc{e}}<1}$ and
	$\tfrac{\mathsf{d}!}{m!(\mathsf{d}-m)!}$ 
	if ${mP_{\textsc{e}}=1}$   
}
\end{equation} 
in number.

\textbf{Remark~1:}
Each of our conditions is related to a witness operator of the kind presented in \cite{Wei03}. 
The condition ${p_\mathsf{k}\leq P_\textsc{e}}$ 
[given in \eqref{p<P_E}] is the same as
\begin{equation}
\label{W}
0\leq\langle W_\mathsf{k}\rangle_{\rho}\,,
\quad\mbox{where}\quad
W_\mathsf{k}=P_\textsc{e}\,\mathsf{I}-\varPi_\mathsf{k}
\end{equation}
is the witness operator associated with the projector $\varPi_\mathsf{k}$ [given in \eqref{prob}] and $\mathsf{I}$ is the identity operator on 
$\mathscr{H}_\mathsf{d}$.
Violation of the inequality in \eqref{W}
detects entanglement, and it is violated by ${\rho=|\textsc{e}_\mathsf{k}\rangle\langle\textsc{e}_\mathsf{k}|}$, but not by any of the entangled states ${\rho=|\textsc{e}_\mathsf{k'}\rangle\langle\textsc{e}_\mathsf{k'}|}$,
where ${\mathsf{k'}\neq\mathsf{k}}$.
So $W_\mathsf{k}$ is a witness (optimal out of a family of witnesses)
for ${|\textsc{e}_\mathsf{k}\rangle}$ \cite{Wei03} and not for the other kets in the entangled basis $\mathfrak{B}_\textsc{e}$.
Since all of $\{W_\mathsf{k}\}_{\mathsf{k}=\mathsf{0}}^{\mathsf{d}-\mathsf{1}}$
correspond to mutually exclusive outcomes of the global measurement in $\mathfrak{B}_\textsc{e}$, they can be realized with a single measurement setting. 
Furthermore, in \cite{Wei03},
${1-P_\textsc{e}}$ is presented as the geometric measure of entanglement for ${|\textsc{e}\rangle}$. According to the measure,
smaller $P_\textsc{e}$ indicates that farther ${|\textsc{e}\rangle}$ is from product kets. So it is better to pick ${|\textsc{e}\rangle}$ with as small $P_\textsc{e}$ as possible, therefore we select maximally entangled kets in Sec.~\ref{sec:Examples}.
Furthermore, one can obtain $\mathcal{S}_\textsc{e}$ with smaller size as its extreme points shrink toward its center ${(\tfrac{1}{\mathsf{d}},\cdots,\tfrac{1}{\mathsf{d}})}$ with a better choice of ${|\textsc{e}\rangle}$, and thus one can detect a bigger set of entangled states with the global measurement described by \eqref{Ent-basis}.

\textbf{Remark~2:}
Since the maximum in \eqref{P_E} is taken over all product kets, our conditions in~\eqref{p<P_E} differentiate between \emph{fully} separable and entangled states.
They are indifferent to genuine and biseparable entanglement.
If we take the maximum over all biseparable kets then we get ${P_\textsc{e}^{\text{bi}}\,(\geq P_\textsc{e})}$.
By replacing $P_\textsc{e}$ by $P_\textsc{e}^{\text{bi}}$ in \eqref{p<P_E}, one can immediately distinguish genuine from biseparable entanglement.
Likewise, one can identify triseparable entanglement and so on.
In \cite{Bourennane04}, a method to compute $P_\textsc{e}^{\text{bi}}$ is given, and the witness operators---of the form of \eqref{W}---that
detect genuine rather than biseparable entanglement are developed in \cite{Bourennane04,Acin01,Toth05,Toth05b,Toth07}.

\textbf{Remark~3:} By defining a measure of uncertainty on the probability space $\Omega$, such as 
\begin{equation}
\label{u, H}
u(\vec{p}\,):=\textstyle\sum\nolimits_{\mathsf{k=0}}^{\mathsf{d-1}}
\sqrt{p_\mathsf{k}}
\quad\mbox{or}\quad
h(\vec{p}\,)=-\textstyle\sum\nolimits_{\mathsf{k=0}}^{\mathsf{d-1}}\,
p_\mathsf{k}\log p_\mathsf{k}\qquad
\end{equation}
given in \cite{Sehrawat16},
we can rephrase the detection criterion~\eqref{criterion} as follows.
Since both $u$ and $h$ (that is entropy) are concave functions of $\vec{p}$ and
$\mathcal{S}_\textsc{e}$ is a compact convex set, 
the absolute minima of these functions in $\mathcal{S}_\textsc{e}$ will be at its extreme points such as~\eqref{ext-pt}.
If ${\vec{p}\notin\mathcal{S}_\textsc{e}}$---which diagnoses entanglement by the test~\eqref{criterion}---then the inequalities
\begin{equation}
\label{u< , H< }
u(\vec{e}_m)\leq u(\vec{p}\,)
\quad\mbox{as well as}\quad
h(\vec{e}_m)\leq h(\vec{p}\,)
\end{equation}
will not hold.
So one can say that violation of these inequalities provides sufficient evidence for entanglement.
There are detection methods based on uncertainty measures \cite{Horodecki96b,Hofmann03,Guhne04,Guhne04b,Giovannetti04}.

\textbf{Remark~4:}
Once entanglement is detected by the postulate~\eqref{criterion}, then one can employ
\begin{equation}
\label{e-estimate}
\varepsilon\,(\vec{p},\mathcal{S}_\textsc{e}):=
\inf\big\{D(\vec{p},\vec{s}\,)\,|\,\vec{s}\in\mathcal{S}_\textsc{e}\big\}
\end{equation}
to compute an \emph{amount of violation}, which gives us an estimate of entanglement.
Basically, the function $\varepsilon$
measures a distance between a point ${\vec{p}\in\Omega}$ and the set $\mathcal{S}_\textsc{e}$ by taking, say, the Euclidean metric 
\begin{equation}
\label{E-distance}
D(\vec{p},\vec{s}\,)=
\sqrt{\textstyle\sum\nolimits_{\mathsf{k}=\mathsf{0}}^{\mathsf{d-1}}
	{(p_\mathsf{k}-s_\mathsf{k})}^2}.
\end{equation}
Obviously ${\varepsilon(\vec{p},\mathcal{S}_\textsc{e})=0}$ for every
separable state, because then $\vec{p}$ belongs to $\mathcal{S}_\textsc{e}$.
Note that the estimate given by $\varepsilon$ depends on the entangled ket, and it can get finer with a better choice of ${|\textsc{e}\rangle}$.
Without further elaborating on this, let us move next.

In certain experiments, it is convenient to perform projective measurements on individual subsystems rather than the joint measurement described by an entangled basis $\mathfrak{B}_\textsc{e}$.
For such a situation, the above detection scheme translates as follows.
All the essential details---and citations regarding the Pauli and Clifford operators and about local MUBs---that are used in the remainder of this section
can be found in Appendix~\ref{app:Pauli Op}.

The product-Pauli operators 
\begin{equation}
\label{pro-Pauli-op}
\Lambda^{(\mathsf{x},\mathsf{z})}:= 
X_{\scriptscriptstyle 1}^{x_1}
Z_{\scriptscriptstyle 1}^{z_1}\otimes\cdots\otimes 
X_{\scriptscriptstyle N}^{x_N}
Z_{\scriptscriptstyle N}^{z_N}\
\end{equation}
configure the product-Pauli basis~\eqref{HS-basis N} of the Hilbert-Schmidt space $\mathscr{B}(\mathscr{H}_\mathsf{d})$,
where 
${\mathsf{x}=(x_{\scriptscriptstyle 1},\cdots, x_{\scriptscriptstyle N})
	\in\mathcal{Z}_\mathsf{d}}$   
and likewise ${\mathsf{z}\in\mathcal{Z}_\mathsf{d}}$ [see \eqref{Z_d} and \eqref{N-tuple}].
Hence every operator $O$ on $\mathscr{H}_\mathsf{d}$ can be uniquely expressed as 
\begin{equation}
\label{O-expansion}
O=\textstyle\sum\limits_{\mathsf{x,z\,}\in \mathcal{Z}_\mathsf{d}}
o_\mathsf{x,z}\,\Lambda^{(\mathsf{x},\mathsf{z})}\,,
\ \ \mbox{where}\ \ 
o_\mathsf{x,z}=\tfrac{1}{\mathsf{d}}
\lgroup\Lambda^{(\mathsf{x},\mathsf{z})},O\rgroup_\textsc{hs}
\end{equation}
are $\mathsf{d}^2$ complex numbers.
Through \eqref{pro-Pauli-op}, one can view $\Lambda^{(\mathsf{x},\mathsf{z})}$ as an elementary-correlation operator, and every $O$ is their linear combination \eqref{O-expansion}.
Next, for distinct ${\mathsf{a},\mathsf{b}\in\mathcal{Z}_\mathsf{d}}$, one can build $\mathsf{d}^2$ new operators
\begin{eqnarray}
\label{O-a,b cong}
O^{(\mathsf{a},\mathsf{b})}&:=&
\Lambda^{(\mathsf{a},\mathsf{b})}\,O\,
{\Lambda^{(\mathsf{a},\mathsf{b})}}^\dagger
\\
\label{O-a,b-exp}
&=&\textstyle\sum\limits_{\mathsf{x,z\,}\in \mathcal{Z}_\mathsf{d}}
o_\mathsf{x,z}\,
\Big(\textstyle\prod\nolimits_{i=1}^{N}
\omega_{d_i}^{\,x_i\stackrel{d_i}{\scriptscriptstyle\boxtimes}b_i
	-\,a_i\stackrel{d_i}{\scriptscriptstyle\boxtimes}z_i}\Big)
\Lambda^{(\mathsf{x},\mathsf{z})}\qquad
\end{eqnarray}
via the conjugation relations~\eqref{Conjugation-relation-N}, ${\omega_{d_i}=\exp(\text{i}\tfrac{2\pi}{d_i})}$ and 
${\text{i}=\sqrt{-1}}$.
Obviously, $O^{(\mathsf{a},\mathsf{b})}$ are locally equivalent to the original operator $O$, and some of these could be the same.

Many criteria for entanglement detection---reviewed in \cite{Guhne09} as well as our criterion~\eqref{criterion}---are established by certain inequalities with expectation values of observables (represented by Hermitian operators). If ${O=O^\dagger}$ is a Hermitian operator, so does every $O^{(\mathsf{a},\mathsf{b})}$.
Moreover, every expectation value
\begin{equation}
\label{<O-a,b>}
\big\langle O^{(\mathsf{a},\mathsf{b})}\big\rangle_\rho
=\textstyle\sum\limits_{\mathsf{x,z\,}\in \mathcal{Z}_\mathsf{d}}
o_\mathsf{x,z}
\Big(\textstyle\prod\nolimits_{i=1}^{N}
\omega_{d_i}^{\,x_i\stackrel{d_i}{\scriptscriptstyle\boxtimes}b_i
	-\,a_i\stackrel{d_i}{\scriptscriptstyle\boxtimes}z_i}\Big)
\big\langle\Lambda^{(\mathsf{x},\mathsf{z})}\big\rangle_\rho\,,
\end{equation}
obtained by exploiting Eqs.~\eqref{Exp-O} and the linearity of trace, materializes as a \emph{correlation function}.
Exactly the same set of Pauli operators $\Lambda^{(\mathsf{x},\mathsf{z})}$, that
appear (for which ${o_\mathsf{x,z}\neq0}$) in the decomposition~\eqref{O-expansion} of $O$,
emerges in the resolution~\eqref{O-a,b-exp} of  ${O^{(\mathsf{a},\mathsf{b})}}$ [for example, see Sec.~\ref{sec:Examples}]. 
So we only need to estimate the expectation values for the set to get
${\langle O^{(\mathsf{a},\mathsf{b})}\rangle_\rho}$ for any
$(\mathsf{a},\mathsf{b})$.

Every $\Lambda^{(\mathsf{x},\mathsf{z})}$
has a product eigenbasis, which is made of local MUBs and totally describes a single local-measurement setting [for a better understanding, see the two paragraphs carrying \eqref{pro basis} and \eqref{x'+z=x+z' N}]. 
Moreover, if and only if $\Lambda^{(\mathsf{x},\mathsf{z})}$ and $\Lambda^{(\mathsf{x}',\mathsf{z}')}$ fulfill all the requirements stated in \eqref{comm-power N}, then their expectation values
can be estimated in a single setting.
We invoke result~\eqref{comm-power N} for counting the number of distinct settings required to estimate ${\langle O^{(\mathsf{a},\mathsf{b})}\rangle_\rho}$, and every setting is laid out as \eqref{pro basis}.
In total, there are ${\textstyle\prod\nolimits_{i=1}^{N} (d_i+1)}$
distinct local-MUB combinations (thus, settings), that will serve all the purposes.

Let us now turn to our projectors given in \eqref{prob}.
If ${O=\varPi=|\textsc{e}\rangle\langle\textsc{e}|}$, then every
$O^{(\mathsf{a,b})}$ is also a projector.
Furthermore, the restrictions~\eqref{p-const2} and \eqref{p<P_E} on probability ${p=\langle\varPi\rangle_\rho}$ turn into
limitations on the associated $N$-party correlation functions:
\begin{equation}
\label{0<varPi-expt<PE}
0\leq\textstyle\sum\limits_{\mathsf{x,z\,}\in \mathcal{Z}_\mathsf{d}}
o_\mathsf{x,z}
\Big(\textstyle\prod\nolimits_{i=1}^{N}
\omega_{d_i}^{\,x_i\stackrel{d_i}{\scriptscriptstyle\boxtimes}b_i
	-\,a_i\stackrel{d_i}{\scriptscriptstyle\boxtimes}z_i}\Big)
\big\langle\Lambda^{(\mathsf{x},\mathsf{z})}\big\rangle_\rho\leq P_\textsc{e}\,.
\end{equation}
Every state $\rho$ respects the left-hand side inequality, which can be used for checking whether an output provided by a computer program [such as developed in Sec.~\ref{subsec:2-qudit}] is correct or not. While violation of the right-hand side inequality for any ${(\mathsf{a,b})}$ detects entanglement.

Here one can recognize that $P_\textsc{e}$ acts as an upper bound on the amount of correlation, and clearly every entangled state made of a
ket of the basis~\eqref{Ent-basis} violates some of these inequalities.
It reveals---quantum correlations are stronger than the classical ones---a common attribute of quantum entanglement.
There are entanglement detection schemes based on certain correlation functions \cite{Kothe07,Maccone15}.
In our case, we do not choose correlation functions before hand---but the entangled projector $\varPi$---they come naturally from the resolution~\eqref{O-expansion} and conjugations~\eqref{O-a,b cong}.

Now one can also acknowledge that the condition ${p\leq P_\textsc{e}}$ is associated with the projector ${|\textsc{e}\rangle\langle\textsc{e}|}$ not with the basis $\mathfrak{B}_\textsc{e}$, thus one can exploit it even if
${|\textsc{e}\rangle}$ belongs to an arbitrary orthonormal basis of 
$\mathscr{H}_\mathsf{d}$.
Besides, one can adopt local, rather than the global, measurements for the detection as described above.
Nevertheless, $\mathfrak{B}_\textsc{e}$, equivalently $\mathcal{L}_\textsc{e}$ [see \eqref{local}], exists for every entangled ket presented in the next section.
In each case, $\mathcal{L}_\textsc{e}$ is contained in the product-Pauli group $\mathcal{P}_\mathsf{d}$, that is every $\mathsf{L_k}$ is some $\Lambda^{(\mathsf{a,b})}$, and thus $\varPi_\mathsf{k}=\varPi^{(\mathsf{a,b})}$.

\textbf{Remark~5:} 
Like the separable set $\mathcal{S}_\textsc{e}$ is bounded by $2\mathsf{d}$ hyperplanes in $\mathbb{R}^{\mathsf{d}}$ [see the text around \eqref{ext-pt}], every equality in \eqref{0<varPi-expt<PE} defines a hyperplane in the $\mathsf{d}^2$-dimensional operator space $\mathscr{B}(\mathscr{H}_\mathsf{d})$.
To comprehend this, let us take another viewpoint.
The Hilbert-Schmidt inner product establishes a one-to-one correspondence between $\mathscr{B}(\mathscr{H}_\mathsf{d})$ and the $\mathsf{d}^2$-dimensional complex-inner-product space
$\mathbb{C}^{\mathsf{d}^2}$ via the right-hand side Eq.~\eqref{O-expansion}.
So we can fully characterize state $\rho$ for our composite system by a unique vector $\overrightarrow{\rho}$ of $\mathsf{d}^2$ complex numbers $r_\mathsf{x,z}$: 
\begin{equation}
	\label{rho-expansion}
\rho=\textstyle\sum\limits_{\mathsf{x,z\,}\in \mathcal{Z}_\mathsf{d}}
	r_\mathsf{x,z}\,\Lambda^{(\mathsf{x},\mathsf{z})}\,,\quad
	r_\mathsf{x,z}=\tfrac{\lgroup\Lambda^{(\mathsf{x},\mathsf{z})},\,\rho\rgroup_\textsc{hs}}{\mathsf{d}}
	=\frac{\overline{\langle\Lambda^{(\mathsf{x},\mathsf{z})}\rangle}_\rho}{\mathsf{d}}\,.
\end{equation}
Now, for example  ${\mathsf{(a,b)=(0,0)}}$, inequalities~\eqref{0<varPi-expt<PE} turn into 
\begin{equation}
	\label{h-plane}
0\,\leq\,
\underbrace{\mathsf{d}\bigg(\textstyle\sum\limits_{\mathsf{x,z\,}
\in\mathcal{Z}_\mathsf{d}}\overline{r_\mathsf{x,z}}\,o_\mathsf{x,z}
\bigg)}_{\lgroup\rho,O\rgroup_\textsc{hs}}
\,\leq\, P_{\textsc{e}}\,,
\end{equation}
where the summation in parentheses represents the standard inner product between complex vectors ${\overrightarrow{\rho},\overrightarrow{O}\in\mathbb{C}^{\mathsf{d}^2}}$ (like $\overrightarrow{\rho}$, $\overrightarrow{O}$ is made of $o_\mathsf{x,z}$).
Now one can appreciate that each equality in \eqref{h-plane} defines a hyperplane in ${\mathbb{C}^{\mathsf{d}^2}}$ that is isomorphic to 
$\mathscr{B}(\mathscr{H}_\mathsf{d})$.
Note that all the information about \emph{preparation} of the composite system goes into the statistical operator $\rho$ (hence into $\overrightarrow{\rho}$) and 
all the information about \emph{measurement settings} goes into the
operator $O$ (thus into $\overrightarrow{O}$).

\textbf{Remark~6:}
The product-Pauli operators $\Lambda^{(\mathsf{x},\mathsf{z})}$ are not
the only ones that constitute an orthonormal basis of $\mathscr{B}(\mathscr{H}_\mathsf{d})$.
At own convenience, one can select another operator basis, such as presented in \cite{Yu05},
and can recast every detection technique given in this write-up accordingly.

\textbf{Remark~7:}
Like $W$ in \eqref{W}, suppose $O$ is an entanglement witness operator \cite{Horodecki96,Terhal00,Lewenstein00,Bruss02}, then the violation of
\begin{equation}
\label{witness-cond}
0\leq\langle O\rangle_\rho
\end{equation}
leads to entanglement detection.
Like Remark~5, the equality in \eqref{witness-cond}
also specifies a hyperplane \cite{Bruss02}.
The state $\varrho$ which gives
${\langle O\rangle_\varrho<0}$ is entangled and identified by the witness $O$. 
One can clearly perceive that $O^{(\mathsf{a,b})}$---obtained with the conjugation~\eqref{O-a,b cong}---is also a witness operator, and it detects entanglement of the state $\varrho^{(\mathsf{a,b})}$ [use \eqref{O-a,b cong} again for describing $\varrho^{(\mathsf{a,b})}$].
Instead of a single condition~\eqref{witness-cond}, now we have one for every $\mathsf{a}$ and $\mathsf{b}$ just like before.
Importantly, each of these conditions can be realized by employing a \emph{same} set of local-measurement settings.
Statements similar to those made next can be issued for witness operators.

We achieve more than one condition through the conjugations~\eqref{O-a,b cong} thanks to operators from the product-Pauli group $\mathcal{P}_\mathsf{d}$.
One can gain even more by---having operators from the product-Clifford group $\mathcal{C}_\mathsf{d}$---the Clifford conjugations. 
$\mathcal{P}_\mathsf{d}$ is a normal subgroup---invariant under the Clifford conjugations---of $\mathcal{C}_\mathsf{d}$.
Truly one can get infinitely many conditions if the complete product-unitary group is considered, and to test all these we only need a finite number [see \eqref{Tot-L-setts}] of local-MUB settings. 
However, in this paper, for each subsystem labeled by ${i\in\{1,\cdots,N\}}$, 
we are considering compositions of the Clifford operators
$F_i$ and $V_i$ only.
The operators $F$ and $V$ [defined by \eqref{F_i} and \eqref{V}] do not belong to the Pauli group $\texttt{P}_d$ and generate the whole Clifford group ${\texttt{CF}_d}$ for ${d=2}$ (qubit) \cite{Gottesman98} and ${d=3}$ (qutrit) \cite{Gottesman99}.
To celebrate the full potential of Clifford conjugations we proceeded to the next section.

\section{Examples of entangled kets and bases}\label{sec:Examples}

In this section, we are picking some well-known entangled kets to demonstrate our entanglement-detection schemes.
Here, at each occasion, every item is delivered in the same sequence---entangled ket $|\textsc{e}\rangle$, basis $\mathfrak{B}_\textsc{e}$, the maximum overlap $P_\textsc{e}$, extreme point ${\vec{e}_m}$ of ${\mathcal{S}_\textsc{e}}$, decomposition of (every) entangled projector ${\varPi_\mathsf{k}}$ as a linear combination of the product-Pauli operators $\Lambda^{(\mathsf{x},\mathsf{z})}$, detection conditions in terms of the correlation functions, and additional conditions due to the Clifford conjugations---as the
previous section.

\subsection{2-qudit system}\label{subsec:2-qudit}

In this subsection, we take a joint system of ${N=2}$ qudits.
Since each constituent is a $d$-level quantum system (qudit), we
omit the subscripts of $d_i$, then ${\mathsf{d}=d^2}$, and of the
local-Pauli operators $X_i$ and $Z_i$.
Owing to Einstein, Podolsky, Rosen \cite{Einstein35} and Bohm \cite{Bohm51}, we have a maximally-entangled Bell-ket
\begin{equation}
\label{B-ket}
|\textsc{b}\rangle=\tfrac{1}{\sqrt{d}}
\textstyle\sum\nolimits_{j=0}^{d-1}\,
|j\rangle\otimes|j\rangle
\end{equation}
for the combined system \cite{note}.
Then an orthonormal Bell-basis $\mathfrak{B}_\textsc{b}$ 
of $\mathscr{H}_\mathsf{d}$ (see in \cite{Bennett93,Lawrence02}) is simply a collection of
\begin{equation}
\label{B-basis}
|\textsc{b}_\mathsf{k}\rangle :=
Z^{k_1}\otimes X^{k_2}\,|\textsc{b}\rangle
\quad\mbox{for all}\quad\mathsf{k}=(k_1,k_2)\in\mathcal{Z}_\mathsf{d}
\end{equation}
[see \eqref{N-tuple} for $\mathcal{Z}_\mathsf{d}$].
In the earlier section, $\mathsf{k}$ is taken as an index that runs between $\mathsf{0}$ and ${\mathsf{d-1}}$, here as well in the later parts it is represented by a $N$-tuple of $\mathcal{Z}_\mathsf{d}$.

One can also acquire a Bell-basis by applying the unitary transformation
\begin{equation}
\label{bell-U}
\textbf{B}:=\big(\textstyle\sum\nolimits_{l=0}^{d-1}
|l\rangle\langle l|\otimes X^l\,\big)
\big(F\otimes I\big)
\end{equation}
to the product basis $\mathcal{B}_0$ of \eqref{pro-basis}, see again \cite{note} in this regards.
Observe that \textbf{B} is composed of two unitary transformations: in the right-hand side transformation, $F$ and $I$ are the Fourier operator of Eq.~\eqref{F_i}---that creates an equal superposition of all the kets of basis~\eqref{B_i}---and the identity operator associated with the second qudit, respectively.
The left-hand side transformation then generates the entanglement across subsystems. Due to which, operator \textbf{B} and ket ${|\textsc{b}_\mathsf{k}\rangle}$ cannot be factorized as the product operator $\mathsf{L}$ and ket ${|\Psi\rangle}$ in \eqref{pro}.
By the way, \textbf{B} is a \emph{non-product} Clifford operator \cite{Gottesman98,Gottesman99}, so does its inverse, which transforms a Bell-basis back to the product basis $\mathcal{B}_0$. 
Non-product operators can change mass (as just shown) and class of entanglement, thus we are not considering these to get conditions for the detection.

Here the joint system only has one (bi)partition, and every Schmidt coefficient of ${|\textsc{b}\rangle}$ is $\tfrac{1}{\sqrt{d}}$, so ${P_\textsc{b}=\tfrac{1}{d}}$ according to \cite{Bourennane04}.
In agreement with our criterion~\eqref{criterion}, violation of the right-hand side inequality~\eqref{B-detect-cond} for any $\mathsf{k}$ exposes entanglement among the qudits. 
Next, one can immediately see ${\vec{e}_{\scriptscriptstyle d-1}=
	\big(\tfrac{1}{d},\cdots,\tfrac{1}{d},0,\cdots,0\big)}$
is an extreme point---others are obtained by permuting its coordinates---of the separable set $\mathcal{S}_\textsc{b}$ for a Bell-basis.
These points are ${\tfrac{d^2!}{d!(d^2-d)!}}$ in total.

First we represent the Bell-projector in the basis~\eqref{Basis-2} as
${|\textsc{b}\rangle\langle\textsc{b}|=
\tfrac{1}{d}\textstyle\sum\nolimits_{j,k=0}^{d-1}\,
(|j\rangle\langle k|)^{{\scriptscriptstyle\otimes\,2}}}$.
Then by employing the transformations~\eqref{proj-op} and ${\textstyle\sum\nolimits_{j=0}^{d-1}
\omega_d^{-(z\,\stackrel{d}{\scriptscriptstyle\boxplus}\,z')\,
	\stackrel{d}{\scriptscriptstyle\boxtimes}\,j}=d\,\delta_{z',-z}}$ 
we obtain
\begin{eqnarray}
\label{B-proj-Pauli}
|\textsc{b}\rangle\langle\textsc{b}|&=&
\tfrac{1}{d^2}\textstyle\sum\nolimits_{x,z=0}^{d-1}
X^xZ^z\otimes X^xZ^{-z},\quad\mbox{and}\qquad\\
\label{B-proj-k-Pauli}
\varPi_\mathsf{k}=
|\textsc{b}_\mathsf{k}\rangle\langle\textsc{b}_\mathsf{k}|&=&
Z^{k_1}\otimes X^{k_2}\,
|\textsc{b}\rangle\langle\textsc{b}|\,
Z^{-k_1}\otimes X^{-k_2}.\qquad
\end{eqnarray}
Now we can state ${\mathsf{d}=d^2}$ conditions, one for each $\mathsf{k}\in\mathcal{Z}_\mathsf{d}$, 
\begin{equation}
\label{B-detect-cond}
0\leq
\underbrace{\tfrac{1}{d^2}\textstyle\sum\nolimits_{x,z=0}^{d-1}
\omega_d^{\,x\,\stackrel{d}{\scriptscriptstyle\boxtimes}\,k_1+z\,\stackrel{d}{\scriptscriptstyle\boxtimes}\,k_2}
\langle X^xZ^z\otimes X^xZ^{-z}\rangle_\rho}%
_{\textstyle p_\mathsf{k}\,=\,\langle\varPi_\mathsf{k}\rangle_\rho}
 \leq\underbrace{\tfrac{1}{d}}_{\textstyle P_\textsc{b}}
\end{equation}
for the entanglement detection [compare these with their general form \eqref{0<varPi-expt<PE}].

The expression sandwiched between two inequality signs is obtained from \eqref{B-proj-k-Pauli} with the aid of
conjugation relations~\eqref{Conjugation-relation-1} and \eqref{Conjugation-relation-N}.  
Moreover, it can be evaluated either by using a single global-measurement in the Bell-basis $\mathfrak{B}_\textsc{b}$
or by adopting ${d+1}$ local-measurement settings.
With the global measurement, we obtain probabilities $p_\mathsf{k}$, while the local measurements provide the elementary correlations
${\langle X^xZ^z\otimes X^xZ^{-z}\rangle_\rho}$. 
Eventually, we secure $d^2$ distinct correlation functions, one for each $p_\mathsf{k}$, given in the conditions~\eqref{B-detect-cond}.
For a general form of this analysis, the reader can always refer back to Sec.~\ref{sec:criteria}.

To understand that we need not more than ${d+1}$ local settings, let us recall from Appendix~\ref{app:composit Pauli Gp} that a single setting is categorized by assigning a single-qudit MUB~\eqref{d+1 bases} to each qudit. 
Subsequently, one can notice that the basis alloted to one qudit completely determines the basis for other, because both the Pauli operators in the tensor product ${X^xZ^z\otimes X^xZ^{-z}}$ are characterized by the same ${(x,z)}$. So we are free to choose a MUB for one qudit only, and since there are ${d+1}$ MUBs~\eqref{d+1 bases} we have ${d+1}$ local settings
\begin{equation}
\label{B-all-loc-settings}
\{0,0\}\,,\,\{1,d-1\}\,,\,\{2,d-2\}\,,\,\cdots\,,\,\{d-1,1\}\,,\,\{d,d\}
\,,
\end{equation}
each of which is displayed like \eqref{pro basis}.

Now, let us discover more conditions for the detection by the virtue of Clifford conjugations. 
For this purpose, 
we study the case of qubit (${d=2}$) and qutrit (${d=3}$) separately in the successive parts.


\subsubsection{2-qubit system}\label{subsubsec:d=2x2}

For a single qubit, the complete Clifford group $\texttt{CF}_2$ can be created by multiplying $F$ and $V$ of Eqs.~\eqref{F_i} and \eqref{V}, respectively \cite{Gottesman98}.
Here we are picking only three compositions
\begin{equation}
\label{Q,T}
 T:=VF\,,\quad T^2\,,\quad\mbox{and}\quad Q:=VFV
\end{equation}
of $F$ and $V$.
None of these five operators is owned by the one-qubit Pauli group $\texttt{P}_2$ [defined by \eqref{Pauli-gp}], and with the identity operator $I$ they grant six automorphisms of $\texttt{P}_2$ under the conjugation \cite{Gottesman98}.
(Note that here $T$ 
is slightly different than it is in \cite{Gottesman98}, but $Q$ is the same.) 
By applying ${T,T^2,Q,F,V}$ to the first-qubit kets, we transform the Bell-basis $\mathfrak{B}_\textsc{b}$ into ${\mathfrak{B}_\textsc{b}^1,\cdots,\mathfrak{B}^5_\textsc{b}}$, respectively.

We can divide these six Bell-bases---that describe half-dozen physically unalike measurements---into two disjoint sets (trios)
\begin{equation}
\label{Bell MUB}
\big\{\mathfrak{B}_\textsc{b}\,,\,\mathfrak{B}_\textsc{b}^1\,,\,
\mathfrak{B}^2_\textsc{b}\big\}
\quad\mbox{and}\quad 
\big\{\mathfrak{B}_\textsc{b}^3\,,\,\mathfrak{B}_\textsc{b}^4\,,\,
\mathfrak{B}^5_\textsc{b}\big\}\,.
\end{equation}
Bases within each trio---but not across the trios---are mutually unbiased (this information is implicitly present in \cite{Klimov07}).
That is, if our two-qubit system is in a Bell-state affiliated to one of the bases, say $\mathfrak{B}_\textsc{b}^1$, and we perform measurement in another basis but from the same trio, say $\mathfrak{B}_\textsc{b}^2$, then each outcome is equally probable (${p_\textsf{k}=\tfrac{1}{4}}$ for every \textsf{k}, which is the center of $\mathcal{S}_\textsc{b}$).
On the other hand, if we perform measurement in a basis from the other trio, say in $\mathfrak{B}_\textsc{b}^3$, then $\vec{p}$ emerges as
an extreme point of $\mathcal{S}_\textsc{b}$.
Note that $\mathcal{S}_\textsc{b}$ associated with $\mathfrak{B}_\textsc{b}^2$ is not the same $\mathcal{S}_\textsc{b}$
with $\mathfrak{B}_\textsc{b}^3$ as the two measurements are
fundamentally different,
although every separable set is defined by \eqref{p-const1}--\eqref{p<P_E}.
In conclusion, every ket of a Bell-basis passes all the tests~\eqref{B-detect-cond} posed by the other bases and remains unseen; it gets detected only by its own basis.
So here we gain total ${6\times4=24}$ distinct conditions (4 for each basis) of the form~\eqref{B-detect-cond}, ${0\leq p_\textsf{k} \leq\tfrac{1}{2}}$, for the detection.
Furthermore, $\mathcal{S}_\textsc{b}$ is an octahedron with six extreme points---${(\tfrac{1}{2},\tfrac{1}{2},0,0)}$ plus five others due to the permutations of coordinates---for every two-qubit Bell-basis.

One must not confuse the Bell-MUB sets~\eqref{Bell MUB} with the local MUBs~\eqref{d+1 bases}, a collection~\eqref{pro basis} of which specifies a local-measurement setting.
For a qudit, every MUB is a common eigenbasis of a set $\texttt{S}^{(x,z)}_d$ [see \eqref{commut-set}] of ${d-1}$ commuting operators, and there are ${d+1}$ such sets [see \eqref{d+1 subsets}].
In the case of qubit (${d=2}$), every \texttt{S} holds a single operator $X$ or ${XZ:=-\text{i}Y}$ or $Z$, so one can alternatively 
represent a local setting with a tensor product 
of the three Pauli operators.
As per the count~\eqref{Tot-L-setts}, there are total 9 local settings for two qubits, which are depicted in Table~\ref{tab:1 d=2} as well as in Table~\ref{tab:2 d=2} by the product operators. 
These two tables also convey the information about decompositions, such as \eqref{B-proj-Pauli}, of the Bell-projectors.

One can check that the original Bell-basis $\mathfrak{B}_\textsc{b}$
is a---unique up to a permutation of and
global phases to its kets---joint eigenbasis of all the ${d^2=4}$ product operators that appear in the resolution~\eqref{B-proj-Pauli}.
One of these is ${I\otimes I}$, and the remaining 3 are
painted with yellow color (and tagged with $\ast$) in Table~\ref{tab:1 d=2}, whose expectation values are evaluated by employing the 3 local settings~\eqref{B-all-loc-settings}.
So the one-to-one relation between the product-operators and the local settings is evident here.

\begin{table}[H]
	\caption{(Color online) The product-Pauli operators with yellow, green, and red shades---also with $\ast$, $\bullet$, and $\circ$ marks---correspond to the Bell-bases $\mathfrak{B}_\textsc{b}$, $\mathfrak{B}_\textsc{b}^1$, and $\mathfrak{B}_\textsc{b}^2$.
	This table with the inequalities \eqref{- sign cond} contributes 12 distinct conditions for detecting the entanglement.}
	\label{tab:1 d=2}
	\centering
	\begin{tabular}{ccc}
		\vspace{1mm} 
		\cellcolor{yellow} ${X\otimes X_{\,\ast}}$ &   
		\cellcolor{green}  ${X\otimes Y_{\,\bullet}}$  & 
		\cellcolor{red}    ${X\otimes Z_{\,\circ}}$  
		\\ 	    \vspace{1mm} 
		\cellcolor{red}    ${Y\otimes X_{\,\circ}}$ &     
		\cellcolor{yellow} ${Y\otimes Y_{\,\ast}}$  & 
		\cellcolor{green}  ${Y\otimes Z_{\,\bullet}}$  
		\\ 	    \vspace{1mm} 
		\cellcolor{green}  ${Z\otimes X_{\,\bullet}}$ &   
		\cellcolor{red}    ${Z\otimes Y_{\,\circ}}$  &   
		\cellcolor{yellow} ${Z\otimes Z_{\,\ast}}$  \\ 
	\end{tabular}
\end{table}

As we obtain $\mathfrak{B}_\textsc{b}^1$ and $\mathfrak{B}_\textsc{b}^2$
from $\mathfrak{B}_\textsc{b}$ with the Clifford operations 
${T\otimes I}$ and ${T^2\otimes I}$, respectively, we attain the green ($\bullet$) and red ($\circ$) colored sets [given in Table~\ref{tab:1 d=2}] from the yellow set via the associated Clifford conjugations.
Operators in green emerge in the resolutions of projectors attached with the Bell-basis $\mathfrak{B}_\textsc{b}^1$, and the same 
alliance the red set and $\mathfrak{B}_\textsc{b}^2$ hold.
Equivalent to \eqref{B-detect-cond}, there is one condition for every projector.
We present all the conditions---now, in form of correlation functions---associated with
$\mathfrak{B}_\textsc{b}$, $\mathfrak{B}_\textsc{b}^1$, and $\mathfrak{B}_\textsc{b}^2$ in a joint manner. 
Since $I\otimes I$ does not change under any unitary conjugation, we have it in every resolution.
Furthermore, $\langle I\otimes I\rangle_\rho=1$ for every $\rho$, thus a simplification leads us to
\begin{eqnarray}
\label{- sign cond}
&&-1\leq+\,\langle A \rangle_\rho+\langle B \rangle_\rho-\langle C \rangle_\rho\leq1\,,\nonumber\\
&&-1\leq+\,\langle A \rangle_\rho-\langle B \rangle_\rho+\langle C \rangle_\rho\leq1\,,\\
&&-1\leq-\,\langle A \rangle_\rho+\langle B \rangle_\rho+\langle C \rangle_\rho\leq1\,,\quad\mbox{and}\nonumber\\
&&-1\leq-\,\langle A \rangle_\rho-\langle B \rangle_\rho-\langle C \rangle_\rho\leq1\,.\nonumber
\end{eqnarray}
By putting 3 operators with a same color code (tag $\ast$ or $\bullet$ or $\circ$) from Table~\ref{tab:1 d=2} at the places of $A$, $B$, and $C$, we have 4 conditions for each Bell-basis mentioned above.
In \cite{Horodecki96b}, authors arrived at the same 4 conditions---which are affiliated to the yellow set of Table~\ref{tab:1 d=2}---from a different path, and constructed a tetrahedron (analogues to the probability space $\Omega$) and a octahedron (parallel to the separable set $\mathcal{S}_\textsc{b}$) with the 4 conditions (see also \cite{Guhne04b}).

\begin{table}[H]
	\caption{(Color online) The product operators colored with yellow, green, and red---and tagged with $\ast$, $\bullet$, and $\circ$---are connected with the Bell-bases $\mathfrak{B}^3_\textsc{b}$, $\mathfrak{B}_\textsc{b}^4$, and $\mathfrak{B}_\textsc{b}^5$, respectively. This table supplies a dozen conditions for the detection through the inequalities \eqref{+ sign cond}.}
	\label{tab:2 d=2}
	\centering
	\begin{tabular}{ccc}
		\vspace{1mm} 
		\cellcolor{yellow} ${X\otimes X_{\,\ast}}$ &   
		\cellcolor{red} ${X\otimes Y_{\,\circ}}$  & 
		\cellcolor{green} ${X\otimes Z_{\,\bullet}}$  
		\\ 	    \vspace{1mm} 
		\cellcolor{red} ${Y\otimes X_{\,\circ}}$ &   
		\cellcolor{green} ${Y\otimes Y_{\,\bullet}}$  & 
		\cellcolor{yellow} ${Y\otimes Z_{\,\ast}}$  
		\\ 	    \vspace{1mm} 
		\cellcolor{green} ${Z\otimes X_{\,\bullet}}$ &   
		\cellcolor{yellow} ${Z\otimes Y_{\,\ast}}$  & 	
		\cellcolor{red} ${Z\otimes Z_{\,\circ}}$  \\ 
	\end{tabular}
\end{table}

Now recall that $\mathfrak{B}_\textsc{b}^3$,  $\mathfrak{B}_\textsc{b}^4$, and $\mathfrak{B}_\textsc{b}^5$
are acquired from the original basis $\mathfrak{B}_\textsc{b}$ by the means of Clifford operations ${Q\otimes I}$, ${F\otimes I}$, and ${V\otimes I}$, correspondingly.
So their conjugations transform the original-yellow set of Table~\ref{tab:1 d=2} into the yellow (tagged with $\ast$),
green ($\bullet$), and red ($\circ$) colored sets of Table~\ref{tab:2 d=2}, in that order.
As before, by placing like-colored operators (with identical tags) from Table~\ref{tab:2 d=2} at the positions of $A$, $B$, and $C$, in
\begin{eqnarray}
\label{+ sign cond}
&&-1\leq-\,\langle A \rangle_\rho-\langle B \rangle_\rho+\langle C \rangle_\rho\leq1\,,\nonumber\\
&&-1\leq-\,\langle A \rangle_\rho+\langle B \rangle_\rho-\langle C \rangle_\rho\leq1\,,\\
&&-1\leq+\,\langle A \rangle_\rho-\langle B \rangle_\rho-\langle C \rangle_\rho\leq1\,,\quad\mbox{and}\nonumber\\
&&-1\leq+\,\langle A \rangle_\rho+\langle B \rangle_\rho+\langle C \rangle_\rho\leq1\,,\nonumber
\end{eqnarray}
we have 4 conditions for each $\mathfrak{B}_\textsc{b}^3$,  $\mathfrak{B}_\textsc{b}^4$, and $\mathfrak{B}_\textsc{b}^5$.

In a nutshell, we explicitly deliver ${6\times4=24}$ conditions, now through the inequalities~\eqref{- sign cond} and \eqref{+ sign cond} with Tables~\ref{tab:1 d=2} and \ref{tab:2 d=2},
for detecting two-qubit entanglement. 
All these conditions can be realized
by the 9 local-MUB settings instead of the 6 global-measurements in
the Bell-bases~\eqref{Bell MUB}.
Basically, each condition can be recovered from \eqref{B-detect-cond} by an appropriate Clifford conjugation.
Every two-qubit state $\rho$ obeys all the left-hand side inequalities given in \eqref{- sign cond} and \eqref{+ sign cond}, because these correspond to probabilities $p_\mathsf{k}$ being nonnegative numbers.
Whereas violation of any right-hand side inequality reveals entanglement.
According to the Peres-Horodecki criterion \cite{Peres96, Horodecki96},
the two-qubit Werner state \cite{Werner89}
\begin{equation}
\label{Werner-state}
\qquad\rho{\scriptstyle(w)}
=\tfrac{1-w}{4}\,I\otimes I+w\,\varPi_{(1,1)}\quad\quad 
(0\leq w\leq 1)\,,
\end{equation}
is entangled if and only if ${\tfrac{1}{3}<w}$, and precisely for all these values of $w$ the last condition of \eqref{- sign cond} 
with the yellow set of Tables~\ref{tab:1 d=2} catches the 
entanglement. 
In state~\eqref{Werner-state}, $\varPi_{(1,1)}$ is the last Bell-projector~\eqref{B-proj-k-Pauli}.

Let us now exploit isomorphism---that is disclosed through Remark~5 in Sec.~\ref{sec:criteria} and is extremely helpful in generating conditions for the detection with an ordinary computer---between the operator space and the complex-vector space.
We associate the four (column) vectors
\begin{equation}
\label{P-vec d=2}
\overrightarrow{I}:=\begin{pmatrix}1 \\0 \\0 \\0 \\\end{pmatrix},\
\overrightarrow{X}:=\begin{pmatrix}0 \\1 \\0 \\0 \\\end{pmatrix},\
\overrightarrow{Y}:=\begin{pmatrix}0 \\0 \\1 \\0 \\\end{pmatrix},\
\overrightarrow{Z}:=\begin{pmatrix}0 \\0 \\0 \\1 \\\end{pmatrix},\
\end{equation}
that constitute the standard basis of $\mathbb{C}^{4}$ to the four operators $I$, $X$, $Y$, and $Z$, respectively, of the Pauli basis~\eqref{d=2 P-basis} of $\mathscr{B}(\mathscr{H}_2)$.
This association establishes a one-to-one correspondence between $\mathscr{B}(\mathscr{H}_2)$ and $\mathbb{C}^{4}$ such that we can uniquely find a vector in $\mathbb{C}^{4}$ for every operator in $\mathscr{B}(\mathscr{H}_2)$ [see Remark~5].

Every unitary conjugation ${A\stackrel{U}{\longrightarrow} UAU^\dagger}$ is a linear map on $\mathscr{B}(\mathscr{H}_2)$, so it can be represented by a ${4\times 4}$ matrix.
Therefore, we have
\begin{equation}
\label{FV-mat d=2}
\widehat{F}=
\begin{pmatrix}
1 & 0 & 0  & 0 \\
0 & 0 & 0  & 1 \\
0 & 0 & -1 & 0 \\
0 & 1 & 0  & 0 \\
\end{pmatrix}\quad\mbox{and}\quad
\widehat{V}=
\begin{pmatrix}
1 & 0 & 0 & 0 \\
0 & 0 & -1 & 0 \\
0 & 1 & 0 & 0 \\
0 & 0 & 0 & 1 \\
\end{pmatrix}\
\end{equation}
that transform vectors~\eqref{P-vec d=2} in the same fashion as 
$F$ and $V$ modify the Pauli basis~\eqref{d=2 P-basis} under the conjugation displayed in Table~\ref{tab:Conj, d=2}.
One can check that $\widehat{F}$ and $\widehat{V}$ follow relations analogues to \eqref{F V cycle} and give $\widehat{I}$.
In fact, one can achieve matrices for every single-qubit Clifford operation by multiplying the above two; for example
\begin{eqnarray}
\label{Q,T mat d=2}
&&\widehat{T}=\widehat{V}\widehat{F},\quad
\widehat{T^2}={\widehat{T}}^{\,2},\quad
\widehat{Q}=
\widehat{V}\widehat{F}\widehat{V},
\quad\mbox{and}\quad\\
\label{ZXY mat d=2}
&&\widehat{Z}={\widehat{V}}^{\,2},\quad
\widehat{X}=\widehat{F}\widehat{Z}\widehat{F},\quad
\widehat{Y}=\widehat{X}\widehat{Z}.
\end{eqnarray} 
Equations~\eqref{Q,T mat d=2} are counterparts of Eqs.~\eqref{Q,T}, and matrices~\eqref{ZXY mat d=2} are for the Pauli operators as they also belong to the Clifford group $\texttt{CF}_2$.
Matrices transform vectors, so we must not mix the roles of matrices~\eqref{ZXY mat d=2} (illustrated with $\widehat{\ }$ ) 
and vectors~\eqref{P-vec d=2}  (exhibited with $\rightarrow$).
Since both ${XZ}$ and ${\text{i}XZ=Y}$ transform---under the conjugation---every single-qubit operator in the same way, we
define matrix $\widehat{Y}$ without the imaginary unit $\text{i}$.

We can always resolve ${O\in\mathscr{B}(\mathscr{H}_\mathsf{d})}$ into a linear combination~\eqref{O-expansion} of the product-Pauli operators.
So, for ${N}$ qubits, ${\mathsf{d}=2^{\scriptscriptstyle N}}$,
by replacing every product operator in \eqref{O-expansion} with the corresponding tensor product of vectors~\eqref{P-vec d=2} we procure
${\overrightarrow{O}\in\mathbb{C}^{\mathsf{d}^2}}$.
Recall from Remark~5 that $\overrightarrow{O}$ is the complex vector associated with an operator $O$.
A tensor product of matrices---such as \eqref{FV-mat d=2}--\eqref{ZXY mat d=2} for each qubit---yields a
${4^{\scriptscriptstyle N}\times4^{\scriptscriptstyle N}}$ matrix, which acts on the $4^{\scriptscriptstyle N}$-component vector $\overrightarrow{O}$ and represents the transformation of operator $O$ under the Clifford conjugation.
Every \emph{distinct} vector obtained in this way specifies a unique condition such as \eqref{h-plane} for the detection.

To generate all the conditions by using a computer program, we need to provide only the vector $\overrightarrow{O}$ and matrices $\widehat{F},\,\widehat{V}$ and $\widehat{I}$ ($=\widehat{F}^{\,4}$) as inputs. 
In the first loop, the program will produce a set of different vectors by applying all possible tensor products of the three matrices on $\overrightarrow{O}$.
In the second loop, it will do the same by taking the generated set---instead of $\overrightarrow{O}$---to achieve a new (and possibly a bigger) set.
After a \emph{finite} number of such feedback loops---feeding the set of dissimilar vectors obtained from the previous loop to the next---we stop getting new additions to the collection.
It is because $F$ and $V$ generate a finite order group. 
Then, to check whether a state $\rho$ for our composite system obeys all these conditions or not, we need to compute the inner products, such as given in parentheses of \eqref{h-plane}, between $\overrightarrow{\rho}$---associated with $\rho$---and vectors of the final collection. 
It is an easy task for a computer.

In case of $N$ qubits, one needs at most $3^{\scriptscriptstyle N}$ [see the tally~\eqref{Tot-L-setts}]
local-measurement settings to test all these conditions experimentally if $\rho$ is unknown.
By the way, one can adopt a similar computer program for the witness operators \cite{Horodecki96,Terhal00,Bruss02,Lewenstein00,Bourennane04}, to draw more than one condition [see Remark~7 in Sec.~\ref{sec:criteria}], as well as when the composite system is made of different species qubits, qutrits, and so on.

After running the above program with the two-qubit Bell-projector~\eqref{B-proj-Pauli} at the place of $O$, we get the same 24 conditions that are already mentioned above.
Now, let us move to the two-qutrit case, where product-Clifford operators provide completely different Bell-bases (thus, conditions) when employed for both the subsystems, unlike here.

\subsubsection{2-qutrit system}\label{subsubsec:d=3x3}

First let us regain from the last paragraph of Sec.~\ref{sec:criteria} 
that $F$ and $V$ [of Eqs.~\eqref{F_i} and \eqref{V}]
create the entire Clifford group $\texttt{CF}_3$ for qutrit \cite{Gottesman99}. 
(With Table~\ref{tab:Conj, d=3}, one can realize that $F^2$ operator provides the necessary $S_a$ gate defined by the mappings (21) and (22) in Ref.~\cite{Gottesman99} for the generation of $\texttt{CF}_3$.)
Record that every local operator in this part of paper acts on single-qutrit kets.

Out of all possible compositions of $F$ and $V$, we select the following
\begin{equation}
	\label{prod of F V}
	\begin{array}%
		{r@{\hspace{4mm}}r@{\hspace{4mm}}r@{\hspace{4mm}}r}
      & F\,,   &VFV^2\,,&V^2FV\,, \vspace{1mm} \\  
V\,,  & VF\,,  &FV\,,   &V^2FV^2\,, \vspace{1mm} \\
V^2\,,& V^2F\,,&FV^2\,, &VFV\,, \\
	\end{array}
\end{equation}
and name these as ${U_1,\cdots,U_{11}}$, from top-left to bottom-right. 
By applying there operators to the first-qutrit kets, we obtain new two-qutrit Bell-bases ${\mathfrak{B}_\textsc{b}^1,\cdots,\mathfrak{B}^{11}_\textsc{b}}$ from the old $\mathfrak{B}_\textsc{b}$, which is the collection of kets~\eqref{B-ket} and \eqref{B-basis}.
Furthermore, if we adopt operators \eqref{prod of F V} for both qutrits instead of one, then we can gain brand new bases (even more conditions for the detection):
the product-Clifford operators 
\begin{equation}
	\label{UxU d=3}
	\begin{array}%
		{l@{\hspace{3mm}}l@{\hspace{3mm}}l@{\hspace{3mm}}l}
		U_1 \otimes U_1\,,& 
		U_2 \otimes U_2\,,&
		U_3 \otimes U_1\,,&
		U_1 \otimes U_3\,, \vspace{1mm} \\  
		U_1 \otimes U_5\,,& 
		U_3 \otimes U_5\,,&
		U_2 \otimes U_7\,,&
		U_1 \otimes U_6\,, \vspace{1mm} \\
		U_1 \otimes U_9\,,& 
		U_1 \otimes U_{11}\,,&
		U_2 \otimes U_{10}\,,&
		U_3 \otimes U_9\,, 
	\end{array}
\end{equation}
transform the original $\mathfrak{B}_\textsc{b}$ into ${\mathfrak{B}_\textsc{b}^{12},\cdots,\mathfrak{B}^{23}_\textsc{b}}$, where the counting is again done from top-left to bottom-right.

Similar to the trios~\eqref{Bell MUB}, here we
segregate the two dozen Bell-bases into three disjoint octets:
\begin{eqnarray}
\label{Bell MUB octet-1}
&&\big\{\mathfrak{B}_\textsc{b}\,,\,\mathfrak{B}^1_\textsc{b}\,,\,
  \mathfrak{B}^2_\textsc{b}\,,\,\mathfrak{B}^3_\textsc{b}\,,\,
  \mathfrak{B}^{12}_\textsc{b}\,,\,\mathfrak{B}^{13}_\textsc{b}\,,\,
  \mathfrak{B}^{14}_\textsc{b}\,,\,\mathfrak{B}^{15}_\textsc{b}\,\big\}
  \,,\quad\\
\label{Bell MUB octet-2}
&&\big\{\mathfrak{B}^{4}_\textsc{b}\,,\,\mathfrak{B}^{5}_\textsc{b}\,,\,
  \mathfrak{B}^{6}_\textsc{b}\,,\,\mathfrak{B}^{7}_\textsc{b}\,,\,
  \mathfrak{B}^{16}_\textsc{b}\,,\,\mathfrak{B}^{17}_\textsc{b}\,,\,
  \mathfrak{B}^{18}_\textsc{b}\,,\,\mathfrak{B}^{19}_\textsc{b}\,\big\}
  \,,\quad \mbox{and}\qquad \ \\
\label{Bell MUB octet-3}
&&\big\{\mathfrak{B}^{8}_\textsc{b}\,,\,\mathfrak{B}^{9}_\textsc{b}\,,\,
  \mathfrak{B}^{10}_\textsc{b}\,,\,\mathfrak{B}^{11}_\textsc{b}\,,\,
  \mathfrak{B}^{20}_\textsc{b}\,,\,\mathfrak{B}^{21}_\textsc{b}\,,\,
  \mathfrak{B}^{22}_\textsc{b}\,,\,\mathfrak{B}^{23}_\textsc{b}\,\big\}
  \,.
\end{eqnarray}
Bases within each octet are mutually unbiased.
Unlike the trios~\eqref{Bell MUB}, 
here a basis of one octet can also be mutually unbiased with a basis---not all the bases---of other octet; for instance, $\mathfrak{B}^1_\textsc{b}$ and $\mathfrak{B}^4_\textsc{b}$.
Therefore, this division~\eqref{Bell MUB octet-1}--\eqref{Bell MUB octet-3} is not unique as that division~\eqref{Bell MUB}.
Nevertheless, the statement---entanglement specified by a ket of a Bell-basis is detected by measurement in the same, not in other, basis---still holds true with respect to these bases: 
if we perform measurement in a different basis, we get a probability vector that either corresponds to the center ${(\tfrac{1}{9},\cdots,\tfrac{1}{9})}$
or to an extreme point such as ${(\tfrac{1}{3},\tfrac{1}{3},\tfrac{1}{3},0,\cdots,0)}$
of the separable set $\mathcal{S}_\textsc{b}$.

Undoubtedly, the two dozen bases portray physically distinct Bell-measurements, and each basis provides 9 different conditions such as \eqref{B-detect-cond}.
All these conditions can be obtained by the Clifford conjugations---of the Bell-projector~\eqref{B-proj-Pauli}---due to the single-qutrit operators~\eqref{prod of F V} and their tensor products \eqref{UxU d=3}.
Moreover, one can check---with a similar computer program that is developed for qubits at the end of Sec.~\ref{subsubsec:d=2x2}---that we get nothing else than the ${24\times9=216}$ conditions in this case with the Clifford conjugations.

For the program, here one needs $9$ vectors
\footnotesize
\begin{equation}
\label{P-vec d=3}
\underbrace{\begin{pmatrix}
	1 \\ 0 \\ 0 \\ 0 \\ 0\\ 0 \\ 0 \\ 0 \\ 0 \\  \end{pmatrix}}_{\overrightarrow{I}},
\underbrace{\begin{pmatrix}
	0 \\ 1 \\ 0 \\ 0 \\ 0\\ 0 \\ 0 \\ 0 \\ 0 \\  \end{pmatrix}}_{\overrightarrow{X}},
\underbrace{\begin{pmatrix}
	0 \\ 0 \\ 1 \\ 0 \\ 0\\ 0 \\ 0 \\ 0 \\ 0 \\  \end{pmatrix}}_{\overrightarrow{X^2}},
\underbrace{\begin{pmatrix}
	0 \\ 0 \\ 0 \\ 1 \\ 0\\ 0 \\ 0 \\ 0 \\ 0 \\  \end{pmatrix}}_{\overrightarrow{XZ}},
\underbrace{\begin{pmatrix}
	0 \\ 0 \\ 0 \\ 0 \\ 1\\ 0 \\ 0 \\ 0 \\ 0 \\  \end{pmatrix}}_{\overrightarrow{X^2Z^2}},
\underbrace{\begin{pmatrix}
	0 \\ 0 \\ 0 \\ 0 \\ 0\\ 1 \\ 0 \\ 0 \\ 0 \\  \end{pmatrix}}_{\overrightarrow{XZ^2}},
\underbrace{\begin{pmatrix}
	0 \\ 0 \\ 0 \\ 0 \\ 0\\ 0 \\ 1 \\ 0 \\ 0 \\  \end{pmatrix}}_{\overrightarrow{X^2Z}},
\underbrace{\begin{pmatrix}
	0 \\ 0 \\ 0 \\ 0 \\ 0\\ 0 \\ 0 \\ 1 \\ 0 \\  \end{pmatrix}}_{\overrightarrow{Z}},
\underbrace{\begin{pmatrix}
	0 \\ 0 \\ 0 \\ 0 \\ 0\\ 0 \\ 0 \\ 0 \\ 1 \\  \end{pmatrix}}_{\overrightarrow{Z^2}}
\end{equation}
\normalsize
of $\mathbb{C}^{9}$ to represent single-qutrit Pauli operators of the basis~\eqref{Basis} of $\mathscr{B}(\mathscr{H}_3)$.
Then, by replacing every operator by its vector in the decomposition~\eqref{B-proj-Pauli}, we have the Bell-projector 
${|\textsc{b}\rangle\langle\textsc{b}|}$ in terms of a complex vector of $9\times 9=81$ components, 9 of which are nonzero.
In the input of program, we provide this vector along with the matrices
\footnotesize
\begin{equation}
\label{FV-mat d=3}
\widehat{F}=
\left(\begin{array}{c|cc|cc|cc|cc}
1 & 0 & 0  & 0 & 0 & 0 & 0  & 0 & 0 \\
\hline
0 & 0 & 0  & 0 & 0 & 0 & 0  & 0 & 1 \\
0 & 0 & 0  & 0 & 0 & 0 & 0  & 1 & 0 \\
\hline
0 & 0 & 0  & 0 & 0 & \omega & 0    & 0 & 0 \\
0 & 0 & 0  & 0 & 0 & 0 & \omega    & 0 & 0 \\
\hline
0 & 0 & 0  & 0 & \omega^2 & 0 & 0  & 0 & 0 \\
0 & 0 & 0  & \omega^2 & 0 & 0 & 0  & 0 & 0 \\
\hline
0 & 1 & 0  & 0 & 0 & 0 & 0  & 0 & 0 \\
0 & 0 & 1  & 0 & 0 & 0 & 0  & 0 & 0 \\
\end{array}\right),
\widehat{V}=
\left(\begin{array}{c|cc|cc|cc|cc}
1 & 0 & 0  & 0 & 0 & 0 & 0  & 0 & 0 \\
\hline
0 & 0 & 0  & 0 & 0 & 1 & 0  & 0 & 0 \\
0 & 0 & 0  & 0 & 0 & 0 & \omega  & 0 & 0 \\
\hline
0 & 1 & 0  & 0 & 0 & 0 & 0  & 0 & 0 \\
0 & 0 & \omega  & 0 & 0 & 0 & 0  & 0 & 0 \\
\hline
0 & 0 & 0  & 1 & 0 & 0 & 0  & 0 & 0 \\
0 & 0 & 0  & 0 & \omega & 0 & 0  & 0 & 0 \\
\hline
0 & 0 & 0  & 0 & 0 & 0 & 0  & 1 & 0 \\
0 & 0 & 0  & 0 & 0 & 0 & 0  & 0 & 1 \\
\end{array}\right),
\end{equation}
\normalsize
and ${\widehat{I}=\widehat{F}^{\,4}}$ to achieve a complete set of distinct vectors through the feedback loops as described in the above section.
In the final set, we secure ${216}$ vectors, each of these holds 9 nonzero entries out of $81$  and presents a unique condition for the detection, like \eqref{B-detect-cond}.
To experimentally test all these conditions, one needs not more than ${4\times4=16}$ 
local-MUB settings [check the number~\eqref{Tot-L-setts}].

\textbf{Remark~8:} For a qubit-qutrit system ${(\mathsf{d}=2\times3)}$, by a similar computer program, we attain ${4896}$ conditions with the Bell-projector onto ${\tfrac{1}{\sqrt{2}}
(|0\rangle\otimes|0\rangle+|1\rangle\otimes|1\rangle)}$,
whose largest Schmidt coefficient is $\tfrac{1}{\sqrt{2}}$. Hence
the maximum overlap is $\tfrac{1}{2}$ \cite{Bourennane04}, which acts as the upper bound in each of these conditions.
One needs ${3\times4=12}$ local-MUB settings to realize all these conditions.
In the input of program, here, we supply a ${36}$-component vector for the Bell-projector, the matrices~\eqref{FV-mat d=2} for qubit, and \eqref{FV-mat d=3} for qutrit.

The matrices~\eqref{FV-mat d=3} are straightforward manifestation of Table~\ref{tab:Conj, d=3}, and one can acquire $\widehat{U}_\alpha$,
${\alpha\in\{1,\cdots,11\}}$, by the composition of $\widehat{F}$ and $\widehat{V}$ analogous to \eqref{prod of F V}. 
The horizontal and vertical lines are placed in matrices \eqref{FV-mat d=3} to clearly visualize how the sets~\eqref{d+1 subsets} and thus the single-qutrit MUBs~\eqref{d+1 bases} get permuted by $F$ and $V$.
More accurately, a local Clifford operator such as $U_\alpha$
transforms a MUB $\texttt{B}_3^t$ into ${\texttt{B}_3^{t'}}$
up to an order of and global phases to the kets.
If we ignore the order and phases, then we can say---indices ${t\in\{0,\cdots,3\}}$ of---local MUBs get permuted by the 
Clifford operators.
After completely removing the first row as well as the first column of $\widehat{F}$,
replacing zero and nonzero blocks (illustrated through the horizontal and vertical lines) by 0 and 1, respectively, we get a ${4\times4}$ matrix that exhibits permutation of the indices.
Likewise, one can have ${4\times4}$ permutation matrix for any $\widehat{U}_\alpha$, and then for ${\widehat{U}_\alpha\otimes\widehat{I}}$ as well as for
${\widehat{U}_\alpha\otimes\widehat{U}_{\alpha'}}$.

\begin{table}[H]
	\caption{(Color online) The original set~\eqref{B-all-loc-settings}---of qutrit-MUB settings associated with $\mathfrak{B}_\textsc{b}$---is dyed here with yellow color and marked by $\ast$. Whereas the blue ($\blacktriangle$), green ($\bullet$), and red ($\circ$) colored (symbolized) sets are related with $\mathfrak{B}^1_\textsc{b}$, $\mathfrak{B}^2_\textsc{b}$, and $\mathfrak{B}^3_\textsc{b}$, respectively.
	Furthermore, these yellow, blue, green, and red sets are also associated
	with $\mathfrak{B}^{12}_\textsc{b}$, $\mathfrak{B}^{13}_\textsc{b}$, $\mathfrak{B}^{14}_\textsc{b}$, and $\mathfrak{B}^{15}_\textsc{b}$, in that order.
	}
	\label{tab:1 d=3}
	\centering
	\begin{tabular}{cccc}
		\vspace{1mm} 
		\cellcolor{yellow} $\{\textbf{0}\,,\,\textbf{0}\}_\ast$ &   
		\cellcolor{green} $\{\textbf{0}\,,\,\textbf{1}\}_\bullet$ & 
		\cellcolor{red} $\{\textbf{0}\,,\,\textbf{2}\}_\circ$ & 
		\cellcolor{blue} $\{\textbf{0}\,,\,\textbf{3}\}_\blacktriangle$    
		\\    \vspace{1mm} 
		\cellcolor{red}    $\{\textbf{1}\,,\,\textbf{0}\}_\circ$ &   
		\cellcolor{blue} $\{\textbf{1}\,,\,\textbf{1}\}_\blacktriangle$ & 
		\cellcolor{yellow} $\{\textbf{1}\,,\,\textbf{2}\}_\ast$ & 
		\cellcolor{green} $\{\textbf{1}\,,\,\textbf{3}\}_\bullet$
		\\    \vspace{1mm}  
		\cellcolor{green}  $\{\textbf{2}\,,\,\textbf{0}\}_\bullet$ &   
		\cellcolor{yellow} $\{\textbf{2}\,,\,\textbf{1}\}_\ast$ & 
		\cellcolor{blue} $\{\textbf{2}\,,\,\textbf{2}\}_\blacktriangle$ & 
		\cellcolor{red} $\{\textbf{2}\,,\,\textbf{3}\}_\circ$  
		\\     \vspace{1mm} 
		\cellcolor{blue} $\{\textbf{3}\,,\,\textbf{0}\}_\blacktriangle$ &   
		\cellcolor{red} $\{\textbf{3}\,,\,\textbf{1}\}_\circ$ & 
		\cellcolor{green} $\{\textbf{3}\,,\,\textbf{2}\}_\bullet$ & 
		\cellcolor{yellow} $\{\textbf{3}\,,\,\textbf{3}\}_\ast$ 
	\end{tabular}
\end{table}

\begin{table}[H]
	\caption{(Color online) Here, the red ($\circ$), blue ($\blacktriangle$), green ($\bullet$), and yellow ($\ast$) painted (tagged) sets of local settings correspond to the pairs 
	${\{\mathfrak{B}^{4}_\textsc{b},\mathfrak{B}^{16}_\textsc{b}\}}$, ${\{\mathfrak{B}^{5}_\textsc{b},\mathfrak{B}^{17}_\textsc{b}\}}$, ${\{\mathfrak{B}^{6}_\textsc{b},\mathfrak{B}^{18}_\textsc{b}\}}$, and 
	${\{\mathfrak{B}^{7}_\textsc{b},\mathfrak{B}^{19}_\textsc{b}\}}$, respectively.}
	\label{tab:2 d=3}
	\centering
	\begin{tabular}{cccc}
		\vspace{1mm} 
		\cellcolor{yellow} $\{\textbf{0}\,,\,\textbf{0}\}_\ast$ &   
		\cellcolor{red} $\{\textbf{0}\,,\,\textbf{1}\}_\circ$ & 
		\cellcolor{blue} $\{\textbf{0}\,,\,\textbf{2}\}_\blacktriangle$ & 
		\cellcolor{green} $\{\textbf{0}\,,\,\textbf{3}\}_\bullet$
		\\    \vspace{1mm}  
		\cellcolor{red} $\{\textbf{1}\,,\,\textbf{0}\}_\circ$ &   
		\cellcolor{yellow} $\{\textbf{1}\,,\,\textbf{1}\}_\ast$ & 
		\cellcolor{green} $\{\textbf{1}\,,\,\textbf{2}\}_\bullet$ & 
		\cellcolor{blue} $\{\textbf{1}\,,\,\textbf{3}\}_\blacktriangle$
		\\     \vspace{1mm} 
		\cellcolor{green} $\{\textbf{2}\,,\,\textbf{0}\}_\bullet$ &   
		\cellcolor{blue} $\{\textbf{2}\,,\,\textbf{1}\}_\blacktriangle$ & 
		\cellcolor{red} $\{\textbf{2}\,,\,\textbf{2}\}_\circ$ & 
		\cellcolor{yellow} $\{\textbf{2}\,,\,\textbf{3}\}_\ast$  
		\\     \vspace{1mm} 
		\cellcolor{blue} $\{\textbf{3}\,,\,\textbf{0}\}_\blacktriangle$ &   
		\cellcolor{green} $\{\textbf{3}\,,\,\textbf{1}\}_\bullet$ & 
		\cellcolor{yellow} $\{\textbf{3}\,,\,\textbf{2}\}_\ast$ & 
		\cellcolor{red} $\{\textbf{3}\,,\,\textbf{3}\}_\circ$  
	\end{tabular}
\end{table}

\begin{table}[H]
	\caption{(Color online) Here, the set with green ($\bullet$), blue ($\blacktriangle$), red ($\circ$), and yellow ($\ast$) shades (signs) are connected with the two-qutrit Bell-bases pairs
	${\{\mathfrak{B}^{8}_\textsc{b},\mathfrak{B}^{20}_\textsc{b}\}}$, ${\{\mathfrak{B}^{9}_\textsc{b},\mathfrak{B}^{21}_\textsc{b}\}}$, ${\{\mathfrak{B}^{10}_\textsc{b},\mathfrak{B}^{22}_\textsc{b}\}}$, and ${\{\mathfrak{B}^{11}_\textsc{b},\mathfrak{B}^{23}_\textsc{b}\}}$, correspondingly.}
	\label{tab:3 d=3}
	\centering
	\begin{tabular}{cccc}
		\vspace{1mm} 
		\cellcolor{yellow} $\{\textbf{0}\,,\,\textbf{0}\}_\ast$ &   
		\cellcolor{blue} $\{\textbf{0}\,,\,\textbf{1}\}_\blacktriangle$ & 
		\cellcolor{green} $\{\textbf{0}\,,\,\textbf{2}\}_\bullet$ &
		\cellcolor{red} $\{\textbf{0}\,,\,\textbf{3}\}_\circ$
		\\     \vspace{1mm}  
		\cellcolor{red} $\{\textbf{1}\,,\,\textbf{0}\}_\circ$ &   
		\cellcolor{green} $\{\textbf{1}\,,\,\textbf{1}\}_\bullet$  & 
		\cellcolor{blue} $\{\textbf{1}\,,\,\textbf{2}\}_\blacktriangle$ & 
		\cellcolor{yellow} $\{\textbf{1}\,,\,\textbf{3}\}_\ast$
		\\      \vspace{1mm} 
		\cellcolor{green} $\{\textbf{2}\,,\,\textbf{0}\}_\bullet$ &   
		\cellcolor{red} $\{\textbf{2}\,,\,\textbf{1}\}_\circ$ & 
		\cellcolor{yellow} $\{\textbf{2}\,,\,\textbf{2}\}_\ast$ & 
		\cellcolor{blue} $\{\textbf{2}\,,\,\textbf{3}\}_\blacktriangle$  
		\\      \vspace{1mm} 
		\cellcolor{blue} $\{\textbf{3}\,,\,\textbf{0}\}_\blacktriangle$ &   
		\cellcolor{yellow} $\{\textbf{3}\,,\,\textbf{1}\}_\ast$ & 
		\cellcolor{red} $\{\textbf{3}\,,\,\textbf{2}\}_\circ$ & 
		\cellcolor{green} $\{\textbf{3}\,,\,\textbf{3}\}_\bullet$  
	\end{tabular}
\end{table}

Recall that the set of ${d+1=4}$ local settings~\eqref{B-all-loc-settings} is associated with the original Bell-basis  $\mathfrak{B}_\textsc{b}$, where
each setting is narrated by a pair of MUB-indices, one for each qutrit.
And, the new Bell-bases are acquired by transforming $\mathfrak{B}_\textsc{b}$ with the single-qutrit Clifford operators~\eqref{prod of F V} and their products~\eqref{UxU d=3}.
Since MUB-indices get permuted by the action of $U_\alpha$, we obtain 12 sets of settings---from the original-set~\eqref{B-all-loc-settings} by the permutation matrices mentioned in above paragraph---for the 24 Bell-bases.
These sets are presented in Tables~\ref{tab:1 d=3}, \ref{tab:2 d=3}, and \ref{tab:3 d=3}, which are associated with the octets~\eqref{Bell MUB octet-1},
\eqref{Bell MUB octet-2}, and \eqref{Bell MUB octet-3}, respectively.
Each table here carries the same 16 local-MUB settings---of course, painted and tagged differently---that will serve all the purposes for a two-qutrit system.

Here one can perceive that each set of settings corresponds to a pair of mutually unbiased Bell-bases, not to a single Bell-basis like Sec.~\ref{subsubsec:d=2x2}.
The sets presented in Tables~\ref{tab:1 d=3}, \ref{tab:2 d=3}, and \ref{tab:3 d=3} do not explicitly carry information about the product-Pauli operators, like Tables~\ref{tab:1 d=2} and \ref{tab:2 d=2}, which emerge in the resolution of a Bell-projector.
Furthermore, every single-qutrit MUB is an eigenbasis of ${d-1=2}$, not 1 like a single-qubit MUB, non-identity linearly-independent Pauli operators [look for $\texttt{S}_d^{(x,z)}$ in Appendix~\ref{app:single Pauli Gp}].
So, here, we do not enjoy one-to-one relation between a product operator and a local-MUB setting.

Now, to check whether a 2-qutrit positive partial transpose (PPT), thus \emph{bound} \cite{Horodecki98}, entangled state is detected by our ${216}$ conditions or not, we consider the state $\varrho_a$ with ${0<a<1}$ given by Eq.~(14) in Ref.~\cite{Horodecki97}.
It is---in the product-basis \cite{note} that we use for the Bell-ket~\eqref{B-ket}---represented by the matrix
\footnotesize
\begin{equation}
\label{bound-ent 3x3}
[\varrho_a]=\tfrac{1}{8a+1}
\left[\begin{array}{ccccccccc}
a & 0 & 0  & 0 & a & 0 & 0  & 0 & a \\
0 & a & 0  & 0 & 0 & 0 & 0  & 0 & 0 \\
0 & 0 & a  & 0 & 0 & 0 & 0  & 0 & 0 \\
0 & 0 & 0  & a & 0 & 0 & 0  & 0 & 0 \\
a & 0 & 0  & 0 & a & 0 & 0  & 0 & a \\
0 & 0 & 0  & 0 & 0 & a & 0  & 0 & 0 \\
0 & 0 & 0  & 0 & 0 & 0 & 
\tfrac{1}{2}(1+a)  & 0 & \tfrac{1}{2}\sqrt{1-a^2}  \\
0 & 0 & 0  & 0 & 0 & 0 & 0  & a & 0 \\
a & 0 & 0  & 0 & a & 0 & 
\tfrac{1}{2}\sqrt{1-a^2}  & 0 & \tfrac{1}{2}(1+a)  \\
\end{array}\right].
\end{equation}
\normalsize
We convert $\varrho_a$ into the vector $\overrightarrow{\varrho_a}$ and compute its inner product with all the ${216}$ vectors in order to apply detection conditions such as~\eqref{h-plane}. 
We have not found any violation for any of the values 
\begin{equation}
\label{set of values}
\left\{\,\tfrac{l}{20}: l=1,2,\cdots,19\,\right\}
\end{equation}
that we tried for $a$.
Hence the bound entanglement of $\varrho_a$ is not detected by the ${216}$ Bell-conditions, at least for this set of values.
It is not surprising because if a state $\rho$
violates condition such as \eqref{B-detect-cond} then it can be distilled \cite{Horodecki99}, then obviously $\rho$ cannot be a bound entangled state.
At the end of next section, it is shown that some 3-qubit PPT entangled states
are detected by the conditions generated by our techniques.


\subsection{\emph{N}-qubit system}\label{subsec:N-qubit}

In this section we review a system of $N$ qubits; so, ${\mathsf{d}=2^{\scriptscriptstyle N}}$. 
Clearly, all the local operators are defined on qubit's Hilbert space $\mathscr{H}_2$ [see Appendix~\ref{app:single Pauli Gp} for their definitions].
For ${N=2}$, every entangled basis presented in the following turns into a 2-qubit Bell-basis stated in Sec.~\ref{subsubsec:d=2x2}.
So naturally a system of ${N\geq3}$ qubits will be our focus here.

Let us begin with the first example \cite{note}: the \textsc{ghz}-state vector \cite{Greenberger89,Greenberger90}
\begin{equation}
\label{ghz-ket}
|\textsc{g}\rangle=\tfrac{1}{\sqrt{2}}
\big[\,|0\rangle^{\scriptscriptstyle\otimes N}+
|1\rangle^{\scriptscriptstyle\otimes N}\,\big]
\end{equation}
for a $N$-qubit system. 
The set of product operators
\begin{equation}
\label{ghz-loc-op-Basis}
\mathsf{L}_\mathsf{k}=Z^{k_1}\otimes X^{k_2}\otimes X^{k_3}\otimes\cdots\otimes X^{k_N}\qquad 
(\mathsf{k}\in\mathcal{Z}_\mathsf{d})
\end{equation}
yields the \textsc{ghz}-basis $\mathfrak{B}_\textsc{g}$ (see in \cite{Guhne04,Guhne04b,Lawrence02}) according to Eqs.~\eqref{local} and \eqref{Ent-basis}.
Here the maximum overlap ${P_\textsc{g}=\tfrac{1}{2}}$ \cite{Wei03} does not depend on $N$, and ${\vec{e}_{\scriptscriptstyle 1}=
(\tfrac{1}{2},\tfrac{1}{2},0,\cdots,0)}$ is an extreme point of the separable set $\mathcal{S}_\textsc{g}$ associated with a \textsc{ghz}-basis.

Parallel to the non-product Clifford operator~\eqref{bell-U}, here we have 
\begin{equation}
\label{ghz-U}
\textbf{G}=\big(
|0\rangle\langle0|\otimes
I^{\scriptscriptstyle\otimes(N-1)}+
|1\rangle\langle 1|\otimes
X^{\scriptscriptstyle\otimes(N-1)}\,\big)
\big(F\otimes I^{\scriptscriptstyle\otimes(N-1)}\big),
\end{equation}
which transforms the product basis $\mathcal{B}_0$ of \eqref{pro-basis} into a \textsc{ghz}-basis \cite{note}. 
Evidently, \textbf{G} is a multiplication of two unitary operations---one creates a superposition and then the other generates the entanglement---just like \textbf{B}.

The \textsc{ghz}-projector
\begin{eqnarray}
\label{|ghz><ghz|}
|\textsc{g}\rangle\langle\textsc{g}|
&=&\tfrac{1}{2}\Big[
\left(\tfrac{I+Z}{2}\right)^{\scriptscriptstyle\otimes N}+
\left(\tfrac{I-Z}{2}\right)^{\scriptscriptstyle\otimes N}+
\nonumber\\
&&\quad \
 \left(\tfrac{X+\text{i}Y}{2}\right)^{\scriptscriptstyle\otimes N}+
\left(\tfrac{X-\text{i}Y}{2}\right)^{\scriptscriptstyle\otimes N}\Big]\quad
\end{eqnarray}
is obtained in this form by the relations~\eqref{proj-op} and \eqref{proj-op N}, and remember that ${Y=\text{i}XZ}$.
Other projectors associated with
$\mathfrak{B}_\textsc{g}$ are ${\varPi_\mathsf{k}=\mathsf{L}_\mathsf{k}|\textsc{g}\rangle\langle\textsc{g}|\mathsf{L}^\dagger_\mathsf{k}}$ as per Eqs.~\eqref{prob} and \eqref{Ent-basis}, moreover ${\mathsf{L}^\dagger_\mathsf{k}=\mathsf{L}_\mathsf{k}}$
[see \eqref{ghz-loc-op-Basis}]. 
Here the criterion~\eqref{criterion} translates as follows: 
if the probability of getting an outcome turns out more than 50\% in a \textsc{ghz}-measurement, such as described by $\mathfrak{B}_\textsc{g}$, then these is entanglement. 
We possess $2^{\scriptscriptstyle N}$ conditions, one for every outcome, of the form~\eqref{p<P_E}.
All of which can be tested either by a single global measurement in 
$\mathfrak{B}_\textsc{g}$, or by employing $1+2^{\scriptscriptstyle N-1}$ local settings.

We count the number of settings by using result~\eqref{comm-power N} in the following manner.
The \textsc{ghz}-projector~\eqref{|ghz><ghz|}---and every other $\varPi_\mathsf{k}$ stated just above---can be expanded further as a linear combination of $2^{\scriptscriptstyle N}$ product-Pauli operators. 
All these operators pairwise commute, but only half of these commute \emph{componentwise}, so we need only 1 local setting for this half.
Since no two operators from the other half commute componentwise, we require $\tfrac{2^N}{2}$ settings, one for each operator, for the second half.

For $N=3$ qubits, not just ${2^{\scriptscriptstyle N}=8}$, but we gain in total ${432}$ disparate conditions for the detection with the help of Clifford conjugations.
We realize this by the computer algorithm presented at the end of
Sec.~\ref{subsubsec:d=2x2}.
By replacing the local Pauli operators
in the projector~\eqref{|ghz><ghz|} with the vectors~\eqref{P-vec d=2}
we own a complex vector of ${4^{\scriptscriptstyle N}=64}$ components---corresponding to ${|\textsc{g}\rangle\langle\textsc{g}|}$---for the input.
Through the feedback mechanism---that requires the $\widehat{F}$ and $\widehat{V}$ matrices of \eqref{FV-mat d=2} and ${\widehat{I}=\widehat{F}^{\,4}}$---we achieve
${432}$ such vectors, each with 8 nonzero entries. 
By the same algorithm, we reach ${2592}$ distinguish conditions with the 4-qubit \textsc{ghz}-projector.
It is not yet clear to us how the number of conditions grows as we increase number $N$ of qubits, which requires further investigation.

For ${N=3}$ qubits, the ${432=54\times8}$ conditions emerge from ${54}$ distinct \textsc{ghz}-bases.
Each of the product-Clifford operators
\begin{equation}
\label{1GHZmub}
\begin{array}%
{l@{\hspace{3mm}}l@{\hspace{3mm}}l}
&
T \otimes T \otimes T\,,& 
T^2 \otimes T^2 \otimes T^2\,, \vspace{1mm} \\  
V \otimes T \otimes T^2\,,&
T^2 \otimes V \otimes T\,,& 
T \otimes T^2 \otimes V\,, \vspace{1mm} 
\end{array}
\end{equation}
transforms the original \textsc{ghz}-basis $\mathfrak{B}_\textsc{g}$
into a locally-equivalent \textsc{ghz}-basis.
Together, with $\mathfrak{B}_\textsc{g}$, they form a set of six  \textsc{ghz}-MUBs; which are presented in \cite{Lawrence02}.
Then, each of the operators
\begin{equation}
\label{5GHZmubs}
I \otimes I \otimes T\,,\,
I \otimes I \otimes T^2\,,\,
I \otimes I \otimes Q\,,\,
I \otimes I \otimes F\,,\,
I \otimes I \otimes V
\end{equation}
converts the MUB-set into a new MUB-set [$T$ and $Q$ are defined in \eqref{Q,T}].
In this way, we have 6 \textsc{ghz}-MUB sets, where each set contains 6 \textsc{ghz}-bases.
In addition, everyone of the Clifford operators 
\begin{equation}
\label{6GHZmub}
\begin{array}%
{l@{\hspace{3mm}}l@{\hspace{3mm}}l}
\big\{\,V   \otimes  T^2 \otimes T^2\,,&
T^2 \otimes  Q   \otimes T\,,&
T   \otimes  I   \otimes V\,\big\}\,,\vspace{1.5mm} \\  
\big\{\,I   \otimes  T^2 \otimes I\,,&
T   \otimes  I   \otimes T\,,&
T^2 \otimes  T   \otimes T^2\,\big\}\,,\vspace{1.5mm} \\  
\big\{\,T   \otimes  V  \otimes T\,,&
T^2 \otimes  T  \otimes T\,,&
I   \otimes  F  \otimes I\,\big\}\,,\vspace{1.5mm} \\  
\big\{\,T   \otimes  V  \otimes V\,,&
T^2 \otimes  Q  \otimes T^2\,,&
V   \otimes  F  \otimes T^2\,\big\}\,,\vspace{1.5mm} \\  
\big\{\,T   \otimes  V  \otimes T^2\,,&
T^2 \otimes  T  \otimes V\,,&
V   \otimes  F  \otimes Q\,\big\}\,,\vspace{1.5mm} \\  
\big\{\,I   \otimes  F  \otimes Q\,,&
T^2 \otimes  Q  \otimes  V\,,&
T   \otimes  V  \otimes F\,\big\}
\end{array}
\end{equation}
transforms the original \textsc{ghz}-basis into a brand new \textsc{ghz}-basis.
In \eqref{6GHZmub}, eighteen operators are divided into 6 sets, in each set three operators provide three mutually unbiased \textsc{ghz}-bases. 
Hence, we obtain ${6\times6+6\times3=54}$ different \textsc{ghz}-bases
and ${54\times8=432}$ locally-equivalent \textsc{ghz}-kets.
The entanglement described by a \textsc{ghz}-ket is detected only by its own projector, not by any of the rest ${431}$ projectors, because the square of the absolute value of the inner product between two distinct \textsc{ghz}-kets is either less than or equal to ${P_\textsc{g}=\tfrac{1}{2}}$.

\textbf{Remark~9:} Both the Bell- and \textsc{ghz}-ket
are examples of (more accurately, locally equivalent to) certain graph-state vectors \cite{Hein06}, whereas our next two examples do not fall into this category. Therefore, all the material presented so far can be directly generalized for the graph kets. 
Witness operators---such as in \eqref{W}---associated with the graph kets are given in \cite{Toth05,Guhne05,Toth05b}.

Our next example of entangled-state vector \cite{note} is the \textsc{w}-ket \cite{Due00}
\begin{equation}
\label{w-ket}
|\textsc{w}\rangle=\tfrac{1}{\sqrt{N}}\,
\textstyle\sum\limits_{\mathfrak{p}}\,
|1\rangle\otimes|0\rangle^{\scriptscriptstyle\otimes\,(N-1)}\,,
\end{equation}
where $\textstyle\sum\nolimits_{\mathfrak{p}}$ stands for the sum over all distinct permutations of items in the tensor product.
It is not known to us how to build a \textsc{w}-basis $\mathfrak{B}_\textsc{w}$---with the \textsc{w}-ket---for an arbitrary number $N$ of qubits.
However, for 3 and 4 qubits, we derive product operators 
\begin{equation}
\label{w-loc-op-Basis}
\mathsf{L}_\mathsf{k}=g_1^{k_1}g_2^{k_2}\cdots g_{\scriptscriptstyle N}^{k_N}\qquad\qquad
(\mathsf{k}\in\mathcal{Z}_\mathsf{d})
\end{equation}
from $g$-operators:
the first and second row
\footnotesize
\begin{eqnarray}
\label{loc-op-g, W-Basis, N=3}
&&
\overbrace{X\otimes I \otimes Z}^{g_1}\quad
\overbrace{Z\otimes X \otimes I}^{g_2}\quad
\overbrace{I\otimes Z \otimes X}^{g_3} \\
\label{loc-op-g, W-Basis, N=4}
&&
\underbrace{Z\otimes Z \otimes I \otimes I}_{g_1}\
\underbrace{Z\otimes I \otimes Z \otimes I}_{g_2}\
\underbrace{X\otimes I \otimes I \otimes I}_{g_3}\
\underbrace{I\otimes X \otimes X \otimes X}_{g_4}\quad \ \ 
\end{eqnarray}
\normalsize 
are reserved to obtain $\mathfrak{B}_\textsc{w}$ for ${N=3}$ and 4 qubits, correspondingly, by applying 
$\mathsf{L}_\mathsf{k}$ of Eq.~\eqref{w-loc-op-Basis} to 
the ket~\eqref{w-ket} [with respect to Eqs.~\eqref{local} and \eqref{Ent-basis}].

It is shown in \cite{Wei03} that the maximum overlap~\eqref{P_E} for every \textsc{w}-ket is
\begin{equation}
P_\textsc{w}=\big(\tfrac{N-1}{N}\big)^{\scriptscriptstyle N-1},
\quad\mbox{and}\quad
\lim_{N\to\infty}P_\textsc{w} =\tfrac{1}{e}\approx0.3678\,,
\end{equation}
where ${e\approx2.718}$ is the Euler's number.
One can see that 
$P_\textsc{w}$ monotonically decreases as $N$ increases. 
For ${N\geq3}$,
${(P_\textsc{w},P_\textsc{w},1-P_\textsc{w},0,\cdots,0)}$ 
presents an extreme point~\eqref{ext-pt} of the separable set $\mathcal{S}_\textsc{w}$ for a \textsc{w}-basis.
At the places of entangling operators~\eqref{bell-U} and \eqref{ghz-U},
here
${\textbf{W}:=
	\textstyle\sum\nolimits_{\mathsf{k}\in\mathcal{Z}_\mathsf{d}}
	|\textsc{w}_\mathsf{k}\rangle\langle\mathsf{k}|}$ 
converts $\mathcal{B}_0$ of \eqref{pro-basis} into the \textsc{w}-basis $\mathfrak{B}_\textsc{w}$ that contains ${|\textsc{w}_\mathsf{k}\rangle=\mathsf{L}_\mathsf{k} |\textsc{w}\rangle}$.

Utilizing the relations~\eqref{proj-op} and \eqref{proj-op N}, we
achieve the following configuration
\begin{eqnarray}
\label{|w><w|}
|\textsc{w}\rangle\langle\textsc{w}|
&=&\tfrac{1}{N}\Big[
\textstyle\sum\limits_{\mathfrak{p}}\tfrac{I-Z}{2}\otimes
\left(\tfrac{I+Z}{2}\right)^{\scriptscriptstyle\otimes\,(N-1)}+
\nonumber\\
&&\quad \ \;  \textstyle\sum\limits_{\mathfrak{p}}
\tfrac{X^{\,\otimes\,2}}{2}\otimes
\left(\tfrac{I+Z}{2}\right)^{\scriptscriptstyle\otimes\,(N-2)}+
\nonumber\\
&&\quad \ \;  \textstyle\sum\limits_{\mathfrak{p}}
\tfrac{Y^{\,\otimes\,2}}{2}\otimes
\left(\tfrac{I+Z}{2}\right)^{\scriptscriptstyle\otimes\,(N-2)}
\;\Big]\quad
\end{eqnarray}
of the \textsc{w}-projector. 
Every product operator in the first summation commutes componentwise with other, not so in the second as well as third summations given above.
So, owing to the result~\eqref{comm-power N}, ${1+2\tfrac{N!}{2!(N-2)!}}$ local settings are essential to compute 
the expectation value of every projector---and to test conditions such as \eqref{p<P_E} for the detection---constructed with $\mathfrak{B}_\textsc{w}$. 
One can easily find these ${1+N(N-1)}$ settings by further expanding \eqref{|w><w|}.

For ${N=3}$, we observe that $\mathfrak{B}_\textsc{w}$ is an eigenbasis of only two linearly-independent product-Pauli operators ${Z\otimes Z\otimes Z}$ and of course ${I\otimes I\otimes I}$.
So applying these to the \textsc{w}-kets do---introduce global phases, but---not deliver anything fresh, whereas the Pauli operators 
\begin{equation}
\label{Pauli-op-w-basis}
{X\otimes I\otimes I},\quad
{Y\otimes I\otimes I},\quad\mbox{and}\quad
{Z\otimes I\otimes I}
\end{equation}
transform $\mathfrak{B}_\textsc{w}$ into completely new \textsc{w}-bases, which are sequentially called $\mathfrak{B}^1_\textsc{w}$, $\mathfrak{B}^2_\textsc{w}$, and $\mathfrak{B}^3_\textsc{w}$.
We have not encountered such a case in the above examples, where 
product-Pauli operators provide only one entangled basis and
$\mathsf{d}$ 
conditions for the detection.
In comparison, here we enjoy ${4\times\mathsf{d}}$ separate conditions thanks to the Pauli operators~\eqref{Pauli-op-w-basis} and 
\eqref{loc-op-g, W-Basis, N=3}.

To comprehend the above paragraph, let us first 
register that the \textsc{ghz}-basis $\mathfrak{B}_\textsc{g}$ is an eigenbasis of ${2^{\scriptscriptstyle N}=8}$ (for ${N=3}$) product-Pauli operators that appear in the expansion~\eqref{|ghz><ghz|}.
These eight with the generating operators~\eqref{ghz-loc-op-Basis} produce, by multiplication, all ${4^{\scriptscriptstyle N}=64}$ members of the product-Pauli basis~\eqref{HS-basis N}.
We are ignoring an overall phase factor to a Pauli operator as it does not have any real consequence here.
Hence Pauli operators do not deliver a new \textsc{ghz}-basis, and the same goes for the Bell-basis $\mathfrak{B}_\textsc{b}$ of Sec.~\ref{subsec:2-qudit}. 
In contrast, here
${Z\otimes Z\otimes Z}$ and the generators ${\{g_1,g_2,g_2\}}$ [given by \eqref{loc-op-g, W-Basis, N=3}] of $\mathfrak{B}_\textsc{w}$ provide only 16, not all ${64}$, operators of 
the Pauli basis~\eqref{HS-basis N}.
None of these sixteen---genuinely changes $\mathfrak{B}_\textsc{w}$---matches with either of the operators~\eqref{Pauli-op-w-basis}. 
So, by multiplying ${X\otimes I\otimes I}$ with the sixteen, we have a new collection of 16 operators.
In this way, with operators~\eqref{Pauli-op-w-basis} and the original sixteen, we have total 4 disjoint sets of Pauli operators, and together they form an operator-basis of ${64}$ elements.
In conclusion, we own one original plus three new \textsc{w}-bases
purely due to the product-Pauli group $\mathcal{P}_\mathsf{d}$.

Bases in the quartet ${\{\mathfrak{B}_\textsc{w},\mathfrak{B}^1_\textsc{w},
\mathfrak{B}^2_\textsc{w},\mathfrak{B}^3_\textsc{w}\}}$ are not mutually unbiased, but (the entanglement represented by) a \textsc{w}-ket is detected by its own basis, not by any other:
a \textsc{w}-ket of one basis corresponds to an extreme point such as $(\tfrac{4}{9},\tfrac{4}{9},\tfrac{1}{9},0,\cdots,0)$ of $\mathcal{S}_\textsc{w}$ that is associated with---measurement in---another basis of the quartet.
It is shown at the end of this section that measurement in $\mathfrak{B}^2_\textsc{w}$ detects PPT entangled states specified by the matrix~\eqref{bound-ent 2x4} for ${b\in(0\,,0.1235]}$.

After including non-Pauli Clifford operators, we achieve ${3456}$ conditions in total for detecting 3-qubit entanglement.
We arrive at this number by the same computer algorithm---laid out at the end of Sec.~\ref{subsubsec:d=2x2}---that we run for the \textsc{ghz}-case.
Here the difference is in the input vector as it is derived from ${|\textsc{w}\rangle\langle\textsc{w}|}$.
Besides, here, every output vector has 20 nonzero components out of 64, since the input vector has so.
It means that the resolution, such as \eqref{|w><w|}, of every 3-qubit \textsc{w}-projector presented here carries only 20 product-Pauli operators.

The hypergraph-state vector \cite{Rossi13,Guhne14}	
\begin{equation}
\label{h-ket}
|\textsc{h}\rangle=
|+\rangle^{\scriptscriptstyle\otimes N}-
\tfrac{2}{\sqrt{2^N}}|1\rangle^{\scriptscriptstyle\otimes N}
\qquad\quad
\big(\,|+\rangle=\tfrac{|0\rangle+|1\rangle}{\sqrt{2}}\,\big)
\end{equation}
is the last example \cite{note} that we are picking in this article.
As described by Eqs.~\eqref{local} and \eqref{Ent-basis}, we achieve the entangled basis $\mathfrak{B}_\textsc{h}$ by the action of
\begin{equation}
\label{h-loc-op-Basis}
\mathsf{L}_\mathsf{k}=Z^{k_1}\otimes Z^{k_2}\otimes\cdots\otimes Z^{k_N}\qquad 
(\mathsf{k}\in\mathcal{Z}_\mathsf{d})
\end{equation}
on the \textsc{h}-ket.
One can also turn the product basis $\mathcal{B}_0$ of \eqref{pro-basis} in to a \textsc{h}-basis with the help of  entangling operator
\begin{eqnarray} 
\label{h-U}
\textbf{H}:=\big(I^{\scriptscriptstyle\,\otimes\, N}-
2\,{(|1\rangle\langle 1|)}^{\scriptscriptstyle\,\otimes\, N}\big)
\big(F^{\scriptscriptstyle\,\otimes\,N}\big)\,.
\end{eqnarray}
Comparable to~\eqref{bell-U} and \eqref{ghz-U}, \textbf{H} is also a composition of two unitary transformations---the first establishes a superposition and the next produces the entanglement.

To compute the maximum overlap~\eqref{P_E} for a \textsc{h}-basis, we take a ket
\begin{equation}
\label{psi_i}
|\psi_i\rangle:=\cos\theta_i\,|0\rangle+
\sin\theta_i\,e^{\text{i}\phi_i}\,|1\rangle
\qquad 
(0\leq \theta_i\,, \tfrac{\phi_i}{4}\leq\tfrac{\pi}{2})
\end{equation}
for every qubit labeled by $i$ and construct their product ket ${|\Psi\rangle}$ according to \eqref{pro}.
Then, the inner product
\begin{equation}
\label{<h|Psi>}
\langle\textsc{h}|\Psi\rangle=
\tfrac{1}{\sqrt{2^N}}\Big[\,
\textstyle\prod\limits_{i=1}^{\scriptscriptstyle N}(\cos\theta_i+\sin\theta_i\,e^{\text{i}\phi_i})-
2\textstyle\prod\limits_{i=1}^{\scriptscriptstyle N}
\sin\theta_i\,e^{\text{i}\phi_i}\,\Big].
\end{equation}
For ${N=2,\cdots,9}$, we numerically compute the maximum $P_\textsc{h}$ of the overlap ${|\langle\textsc{h}|\Psi\rangle|^2}$ (see also \cite{Guhne14,Rossi14}):
\begin{equation}
\begin{array}{c@{\hspace{1.5mm}}|@{\hspace{1.5mm}}c@{\hspace{1.5mm}}c@{\hspace{1.5mm}}c@{\hspace{1.5mm}}c@{\hspace{1.5mm}}c@{\hspace{1.5mm}}c@{\hspace{1.5mm}}c@{\hspace{1.5mm}}c}
N & 2 & 3 & 4 & 5 & 6 & 7 & 8 & 9\\ 
\hline
P_\textsc{h}\, & 
0.50 & 0.67 & 0.80 & 0.89 & 0.94 & 0.97 & 0.98 & 0.99
\end{array}\;.
\end{equation}
Here one can acknowledge that $P_\textsc{h}$ tends to (not equal to) 1 as number of qubits $N$ grows.
It implies---according to the geometric measure ${1-P_\textsc{h}}$ of entanglement \cite{Wei03}---that ${|\textsc{h}\rangle}$ becomes less entangled and more like a product ket for large $N$ \cite{Rossi14}.
Furthermore, ${\vec{e}_{\scriptscriptstyle 1}=(P_\textsc{h}, 1-P_\textsc{h},0,\cdots,0)}$ is an extreme point~\eqref{ext-pt} of the separable set $\mathcal{S}_\textsc{h}$, which for large $N$ almost fills the entire probability space $\Omega$ [defined by \eqref{p-const1} and \eqref{p-const2}].

Let us move to the \textsc{h}-projector
\begin{eqnarray}
\label{|h><h|}
|\textsc{h}\rangle\langle\textsc{h}|
&=&
\big(\tfrac{I+X}{2}\big)^{\scriptscriptstyle\otimes N}+
\tfrac{1}{2^{N-2}}
\big(\tfrac{I-Z}{2}\big)^{\scriptscriptstyle\otimes N}
\nonumber\\
&&-\tfrac{1}{2^{N-1}}\Big[
\big(\tfrac{I+X+\text{i}Y-Z}{2}\big)^{\scriptscriptstyle\otimes N}+
\big(\tfrac{I+X-\text{i}Y-Z}{2}\big)^{\scriptscriptstyle\otimes N}
\Big]\quad\ \;\,
\end{eqnarray}
that we get in the above configuration by the relations~\eqref{proj-op} and \eqref{proj-op N}.
One can recover the other projectors ${\mathsf{L_k}|\textsc{h}\rangle\langle\textsc{h}|\mathsf{L}^\dagger_\mathsf{k}}$ with the product operators~\eqref{h-loc-op-Basis}.
One can
detect $N$-qubit entanglement by global measurement in the \textsc{h}-basis with the criterion~\eqref{criterion} for which $P_\textsc{h}$ is presented above.

As we know, one can employ local MUBs instead, for
the detection [by inequalities~\eqref{0<varPi-expt<PE}].
Unlike the previous examples, it is cumbersome to count the number of local-MUB settings---corresponding to $\mathfrak{B}_\textsc{h}$---for an arbitrary $N$.
Nevertheless, by the principle~\eqref{comm-power N}, we realize that ${13}$ and ${40}$ settings are needed for ${N=3}$ and 4 qubits, in that order.
One can easily recognize these settings by further expansion of \eqref{|h><h|}, which bears ${29}$ and ${121}$ product-Pauli operators in case of ${N=3}$ and 4, respectively.

For ${N=3}$, we notice that $\mathfrak{B}_\textsc{h}$ is eigenbasis of no product-Pauli operator except, of course, scalar multiples of ${I\otimes I\otimes I}$. 
Analogous to the previous example, each of the Pauli operators
\begin{eqnarray}
\label{Pauli-op-h-basis}
&&{X\otimes I\otimes I},\ \,
  {I\otimes X\otimes I},\ \,
  {I\otimes I\otimes X}, \nonumber\\
&&{I\otimes X\otimes X},\ \,
  {X\otimes I\otimes X},\ \,
  {X\otimes X\otimes I},\ \,
  {X\otimes X\otimes X}\qquad
\end{eqnarray}
transforms $\mathfrak{B}_\textsc{h}$ into a new \textsc{h}-basis. For this, one can prepare an explanation similar to that is presented for the \textsc{w}-bases quartet [see the paragraph holding operators~\eqref{Pauli-op-w-basis} and the next].
Not any pair of the \textsc{h}-bases shares a single ket; in fact, the absolute value of inner product between kets belong to different bases is either $\tfrac{1}{2}$ or 0.
Hence these seven plus one bases (octet) specify 8 distinct global measurements, and measurement in one basis does not detect entanglement represented by a ket from another basis, needless to say due to the rule~\eqref{criterion}.
Furthermore, since a Pauli operator only introduces a phase factor to another Pauli operator under the conjugation [see Eqs.~\eqref{Conjugation-relation-1} and \eqref{Conjugation-relation-N}], every \textsc{h}-projector---constructed with a ket from any of the octet---has the same ${29}$ product-Pauli operators in its decomposition.
Consequently, all the eight \textsc{h}-bases (and their ${8\times8}$ conditions) correspond to the \emph{same} 13 local settings.
The matching statement can be issued for the four \textsc{w}-bases presented earlier.

With the computer program---introduced at the end of Sec.~\ref{subsubsec:d=2x2}---we obtain total ${13824}$ conditions here for detecting 3-qubit entanglement.
All these can be realized with
only ${3^{\scriptscriptstyle N}=27}$ local settings [see the total~\eqref{Tot-L-setts}]. 
Essentially, the program picks distinct multiplications of the matrices~\eqref{FV-mat d=2} for each qubit and 
applies their tensor products to the ${64}$-component vector derived from the \textsc{h}-projector~\eqref{|h><h|}.
In the output, ${13824}$ individual vectors, each holds 29 nonzero numbers, are obtained thanks to the Clifford conjugations.

For a comparison, we restate the maximum overlap and the total number of conditions, associated with each of the three examples, for 3-qubit entanglement detection.
\begin{equation}
\label{g w h comparison}
\begin{array}{c@{\hspace{6mm}} c@{\hspace{4mm}}c@{\hspace{4mm}}c}
|\textsc{e}\rangle: & 
|\textsc{g}\rangle & |\textsc{w}\rangle & |\textsc{h}\rangle \\ 
P_\textsc{e}: & 0.5 & 0.444 & 0.676\\
\mbox{Total conditions} : & 432 & 3456 & 13824
\end{array}\;
\end{equation}
If our 3-qubit system is in the \textsc{ghz}-state ${\rho=|\textsc{g}\rangle\langle\textsc{g}|}$, then the entanglement is detected not only by the \textsc{ghz}-conditions but also by the \textsc{w}-conditions. 
Although it is not revealed by the \textsc{h}-conditions.
Whereas, entanglement of ${|\textsc{w}\rangle\langle\textsc{w}|}$ as well as of ${|\textsc{h}\rangle\langle\textsc{h}|}$ are detected by all
the three types of conditions [listed in \eqref{g w h comparison}].
One can easily check all this
as described in third to the last paragraph in Sec.~\ref{subsubsec:d=2x2}.

With the same technique, we inspect that whether entanglement of the state $\sigma_b$ (${0<b<1}$) introduced in \cite{Horodecki97} is detected by our conditions or not.
In the product-basis that we adopt~\cite{note} for the entangled kets, $\sigma_b$ is represented by
\footnotesize
\begin{equation}
\label{bound-ent 2x4}
[\sigma_b]=\tfrac{1}{7b+1}
\left[\begin{array}{cccccccc}
b & 0 & 0  & 0 & 0 & b & 0  & 0  \\
0 & b & 0  & 0 & 0 & 0 & b  & 0  \\
0 & 0 & b  & 0 & 0 & 0 & 0  & b  \\
0 & 0 & 0  & b & 0 & 0 & 0  & 0  \\
0 & 0 & 0  & 0 & 
\tfrac{1}{2}(1+b) &  0 & 0  & \tfrac{1}{2}\sqrt{1-b^2}  \\
b & 0 & 0  & 0 & 0 & b & 0  & 0  \\
0 & b & 0  & 0 & 0 & 0 & b  & 0  \\
0 & 0 & b  & 0 & 
\tfrac{1}{2}\sqrt{1-b^2} & 0 & 0  & \tfrac{1}{2}(1+b)  \\
\end{array}\right].
\end{equation}
\normalsize
There exist three bipartitions of a 3-qubit system:~{1-(23)}, {2-(13)}, and {3-(12)}, where 1, 2, and 3 are labels for the qubits.
$\sigma_b$ is a PPT entangled state \cite{Horodecki97,Horodecki01}
with respect to the bipartition {1-(23)} and not so with respect to the other bipartitions \cite{Thanks}:
${0\leq[\sigma_b]^{T_1}}$, ${0\nleq[\sigma_b]^{T_2}}$, and ${0\nleq[\sigma_b]^{T_3}}$, where $T_i$ stands for 
partial transposition (for the definition, see \cite{Peres96}) of the indices corresponding to $i$th qubit only.
It turns out that our \textsc{ghz}- and \textsc{h}-conditions do not spot the entanglement of $\sigma_b$ for any value drawn from the set~\eqref{set of values} for $b$.
However, the \textsc{w}-conditions reveal the entanglement for $b=\tfrac{1}{8.1},\tfrac{1}{9},\tfrac{1}{10},\tfrac{1}{11}$, and $\tfrac{1}{20}$.
Which can be verify by taking 
${|\textsc{w}'\rangle:=X\otimes X\otimes I|\textsc{w}\rangle}$ and then realizing
\begin{equation}
\langle\textsc{w}\,'|\sigma_b|\textsc{w}\,'\rangle=
\frac{1+4b+\sqrt{1-b^2}}{3\,(7b+1)}>\frac{4}{9}\ \; \mbox{when}\ \; b\leq0.1235\,.
\end{equation}
In fact, the ket ${\text{i}|\textsc{w}'\rangle}$ belongs to the basis
$\mathfrak{B}^2_\textsc{w}$ given above.

\textbf{Remark~10:} For 3-qubit bound entanglement detection, for example, one can consider rank $4$ projector
\begin{eqnarray}
	\label{pro-upb}
	\varPi_{\textsc{upb}}^{\perp}
	=I\otimes I\otimes I&-&
	\Big[\tfrac{I+Z}{2}\otimes\tfrac{I-Z}{2}\otimes\tfrac{I+X}{2}
	\nonumber\\
	&+&\ \; \tfrac{I-Z}{2}\otimes\tfrac{I+X}{2}\otimes\tfrac{I+Z}{2}
	\\
	&+&\ \; \tfrac{I+X}{2}\otimes\tfrac{I+Z}{2}\otimes\tfrac{I-Z}{2}
	\nonumber\\
	&+&\ \; \tfrac{I-X}{2}\otimes\tfrac{I-X}{2}\otimes\tfrac{I-X}{2}
	\;\Big]	\nonumber
\end{eqnarray}
onto the orthogonal complement of the space spanned by the unextendible product basis
\begin{equation}
\label{upb}
\big\{\,|0\rangle|1\rangle|+\rangle\,,\,
|1\rangle|+\rangle|0\rangle\,,\,
|+\rangle|0\rangle|1\rangle\,,\,
|-\rangle|-\rangle|-\rangle\,
\big\}
\end{equation}
given in \cite{Bennett99}. 
In~\eqref{upb}, $\otimes$ is not shown between the kets, and 
${|\pm\rangle=\tfrac{|0\rangle\pm|1\rangle}{\sqrt{2}}}$.
Like $P_\textsc{H}$ above, we numerically compute ${P_\textsc{upb}=
\operatorname*{max}_{|\Psi\rangle}
\langle\Psi|\varPi_{\textsc{upb}}^{\perp}|\Psi\rangle\approx0.9185}$ (see Appendix~A in \cite{Branciard10}).
The projector \eqref{pro-upb} detects entanglement of the 3-qubit state ${\rho_{\textsc{upb}}=\tfrac{1}{4}\varPi_{\textsc{upb}}^{\perp}}$ that violates the condition $\langle\varPi_{\textsc{upb}}^{\perp}\rangle_\rho\leq P_\textsc{upb}$
(for similar approaches, see in \cite{Terhal01,Guhne04}).
In fact, $\rho_{\textsc{upb}}$ is separable with respect to every bipartition \cite{Bennett99,DiVincenzo03} but \emph{not fully} separable (bound entangled \cite{Horodecki01}) according to the range criterion \cite{Horodecki97}.
Furthermore, by taking $\varPi_{\textsc{upb}}^{\perp}$, we obtain ${3456}$ individual conditions through the Clifford conjugations with the computer algorithm presented in Sec.~\ref{subsubsec:d=2x2}.
These conditions also reveal entanglement of  ${|\textsc{g}\rangle}$, ${|\textsc{w}\rangle}$, and ${|\textsc{h}\rangle}$, but not of $\sigma_b$ for any $b$ belongs to the set~\eqref{set of values}.

\section{Conclusion and outlook}\label{sec:Conc}

We present two equivalent schemes for multipartite-entanglement detection.
To apply our schemes, one needs to select an entangled ket ${|\textsc{e}\rangle}$ as far away as possible---determined by the maximum overlap $P_\textsc{e}$---from the product kets. 
By putting $P_\textsc{e}$ as an upper bound on the
expectation value of projector onto the ket, we secure one condition for the detection.
We emphasize that each of our conditions can be expressed with a witness operator such as constructed in \cite{Wei03} (see also witness operators in \cite{Bourennane04,Acin01,Toth05,Guhne04b,Guhne05,Toth05b,Toth07,Terhal01}).
Subsequently, one can either build an entangled basis (such as in \cite{Guhne04,Lawrence02,Klimov07}) with the entangled ket and realize many such conditions with a global projective measurement. 
Or, one can resolve the entangled projector into a linear combination of the product-Pauli operators and test the conditions with local-MUB
measurements.
In addition, one can gain hundreds and thousands of conditions, even for 3 qubits, via the Clifford conjugations.

Since the number is overwhelming, only
for the 2-qubit and 2-qutrit Bell-kets and for the 3-qubit \textsc{ghz}-ket, we present all the conditions explicitly.
For all the (other) cases we provide a powerful yet simple algorithm to obtain every condition and then to check whether a given state obeys all these or not with an ordinary computer.
In fact, one can use our algorithm to gain many conditions from a witness operator, and all those conditions obtained through the product-Pauli operations can be realized by the same set of local-measurement settings.
If state for the compounded system is unknown, no matter how many conditions we have to test experimentally, all we need is
${(d+1)^{\scriptscriptstyle N}}$ local-MUB settings for $N$-qudits.
Although state tomography is not in any way required, but one can do it with the data acquired with these settings.

We also present MUB-structure of a half-dozen 2-qubit and two dozen 2-qutrit
Bell-bases, which specify ${24}$ and ${216}$ conditions, respectively.
It is not clear to us, whether the ${24}$ conditions detect all 2-qubit entangled states or not.
Whereas it is shown above that our conditions do not detect entanglement of every (particularly, bound entangled) state in the 2-qutrit and 3-qubit cases.
We realize that our \textsc{w}-conditions identify some PPT entangled states which remain hidden to the \textsc{ghz}- and \textsc{h}-conditions.
For 3 qubits, MUB-structure of ${54}$ \textsc{ghz}-bases is provided in the paper, which specifies the ${432}$ distinct \textsc{ghz}-conditions.

The bound entanglement can be detected by constructing projectors \cite{Terhal01,Guhne04} using the unextendible product bases.
Choosing different entangled projectors and going beyond the Clifford group---considering unitary operators such as $V^{\frac{1}{2}}$ for the conjugation---could be helpful in detecting a bigger set of (bound) entangled states. These directions need further research.


\begin{acknowledgments} 
I am grateful to Arvind for stimulating discussions, and to Berthold-Georg Englert, Markus Grassl, Gelo Noel M. Tabia (for bringing Ref.~\cite{Gottesman99} to my attention) and Philippe Raynal (for providing Ref.~\cite{Wootters90}) for their helpful correspondences.
\end{acknowledgments}


\appendix
\section{Additive group $\mathcal{Z}_\mathsf{d}$, product-Pauli group $\mathcal{P}_\mathsf{d}$ and product-Clifford group $\mathcal{C}_\mathsf{d}$}\label{app:Pauli Op}

\subsection{For a single subsystem}\label{app:single Pauli Gp}

Most of the material here is borrowed from \cite{Lidl86,Weyl32,Schwinger60,Ivanovic81,Englert01,Bandyopadhyay02,Lawrence02,Klimov05,Klimov07,Durt10,Englert06,Gottesman98,Gottesman99,Wootters90}.
For a prime number $d$, the set of integers
\begin{equation}
\label{Z_d}
\mathbb{Z}_d:=
\big\{\,j\,\big\}_{j=0}^{d-1}
\end{equation}
constitutes a prime field \cite{Lidl86} under 
modulo-$d$ addition and multiplication, which are symbolized by $\stackrel{d}{\scriptstyle\boxplus}$ and $\stackrel{d}{\scriptstyle\boxtimes}$, respectively.

Now suppose
\begin{equation}
\label{B_i}
\qquad
\texttt{B}^{d}_{d}:=
\big\{|j\rangle:j\in\mathbb{Z}_d\big\}
\qquad\qquad
\big(\langle j'|j\rangle=\delta_{j,j'}\big)
\end{equation}
is an arbitrary orthonormal basis of Hilbert space $\mathscr{H}_d$, where the kets are labeled
by the integers of $\mathbb{Z}_d$.
Recall that $\delta_{j,j'}$ is the Kronecker delta function.
Taking $\texttt{B}^d_d$, we construct a pair of unitary operators \cite{Schwinger60,Englert01,Englert06}
\begin{eqnarray}
\label{X_i}
X&:=&\textstyle\sum\nolimits_{j=0}^{d-1}\,
\big|j\stackrel{d}{\scriptstyle\boxplus}1\big\rangle\big\langle j\big| \qquad\quad\qquad(X^d=I)\quad\mbox{and}\qquad\\
\label{Z_i}
Z&:=&\textstyle\sum\nolimits_{j=0}^{d-1}\,
\omega_{d}^{\,j}\,|j\rangle\langle j| 
\qquad\quad\qquad\ \ (Z^d=I)\,,
\end{eqnarray}
where $I$ is the identity operator on $\mathscr{H}_{d}$, 
${\omega_d=\exp(\text{i}\tfrac{2\pi}{d})}$, and
${\text{i}=\sqrt{-1}}$.
These operators share the following relations \cite{Gottesman98,Gottesman99}
\begin{eqnarray}
\label{FXF,FZF}
FXF^\dagger&=&Z\quad\mbox{and}\quad FZF^\dagger=X^{d-1}\,,
\quad\mbox{where}\quad\\
\label{F_i}
F&:=&\tfrac{1}{\sqrt{d}}\textstyle\sum\nolimits_{j,j'=0}^{d-1}
\omega_{d}^{\;j\,\stackrel{d}{\scriptscriptstyle\boxtimes}\,j'}|j'\rangle\langle j|
\end{eqnarray}
is the discrete Fourier transformation (unitary operator).

By multiplying $X$ and $Z$, one can build the generalized Pauli,
also known as Heisenberg-Weyl, group  \cite{Durt10} 
\begin{equation}
   \label{Pauli-gp}
   \texttt{P}_d:=\big\{\omega_d^wX^xZ^z:w,x,z\in\mathbb{Z}_d\big\}
\end{equation}
of $d^3$ elements.
With the Weyl commutation relation \cite{Weyl32}
\begin{equation}
\label{Weyl commutation}
   Z^zX^x=
   \omega_d^{\,x\,\stackrel{d}{\scriptscriptstyle\boxtimes}\,z} X^xZ^z \quad\text{for every}\quad x,z\in\mathbb{Z}_d
\end{equation}
one can perceive that ${Z^zX^x}$ are already included in \eqref{Pauli-gp}.
Since $ZX$-ordered compositions are linearly dependent on (scalar multiplication of) $XZ$-ordered compositions, we consider the subset
\begin{equation}
\label{Basis}
\big\{X^xZ^z:x,z\in\mathbb{Z}_d\big\}
\end{equation}
of $\texttt{P}_d$.
It carries $d^2$ elements and constitutes an orthogonal basis \cite{Schwinger60,Englert01,Bandyopadhyay02} of the
$d^2$-dimensional Hilbert-Schmidt space $\mathscr{B}(\mathscr{H}_d)$
because
\begin{equation}
\label{orth-XZ}
\big\lgroup X^{x'}Z^{z'},X^xZ^z\big\rgroup_\textsc{hs}=
\text{tr}\big(X^{x-x'}Z^{z-z'}\big)=
d\,\delta_{x,x'}\delta_{z,z'}\qquad
\end{equation}
[see definition~\eqref{HS-inner-pro} of the inner product].
The set~\eqref{Basis} is titled as \emph{Pauli basis}.

With the relation~\eqref{Weyl commutation}, one can prove that
two Pauli operators commute \cite{Bandyopadhyay02,Gottesman99}, that is 
\begin{equation}
\label{commute}
(X^{x'}Z^{z'})(X^xZ^z)=(X^xZ^z)(X^{x'}Z^{z'})\,,
\end{equation}
if and only if
\begin{equation}
\label{x'+z=x+z'}
x\stackrel{d}{\scriptstyle\boxtimes}z'=
x'\stackrel{d}{\scriptstyle\boxtimes}z
\quad(\text{mod}\ d)\,.
\end{equation}
To assemble a set of pairwise commuting operators, we need to find all ${(x',z')}$ that respect Eq.~\eqref{x'+z=x+z'} for a given ${(x,z)}$.
Let us consider the cases ${x=0=z}$, ${x\neq0=z}$, ${x=0\neq z}$, and ${x\neq0\neq z}$ one by one.
For ${x=0=z}$, every ${(x',z')}$ satisfies Eq.~\eqref{x'+z=x+z'}, which simply means that every ${X^{x'}Z^{z'}}$ commutes with the identity operator.
\begin{equation}
\label{comm-power}
\parbox{0.7\columnwidth}
{In all other cases, except ${x=0=z}$, an operator ${X^{x'}Z^{z'}}$ commutes with ${X^xZ^z}$ if and only if there exists ${k\in\mathbb{Z}_d}$ such that ${(x',z')=k\stackrel{d}{\scriptstyle\boxtimes}(x,z)}$.
}
\end{equation}

For ${x\neq0=z}$, Eq.~\eqref{x'+z=x+z'} transforms into ${x\stackrel{d}{\scriptstyle\boxtimes}z'=0}$ that is obeyed by every ${x'\in\mathbb{Z}_d}$, but only by ${z'=0}$.
For a nonzero $x$, the set ${\big\{k\stackrel{d}{\scriptstyle\boxtimes}x:k\in\mathbb{Z}_d\big\}}$ is nothing but $\mathbb{Z}_d$ itself, consequently we can find ${k\in\mathbb{Z}_d}$ such that ${x'=k\stackrel{d}{\scriptstyle\boxtimes}x}$. With the same $k$, we can also express ${z'=k\stackrel{d}{\scriptstyle\boxtimes}z=0}$.
The case ${x=0\neq z}$ can be handled in the same fashion.
For ${x\neq0\neq z}$, like above, we can find ${k,l\in\mathbb{Z}_d}$ such that ${x'=k\stackrel{d}{\scriptstyle\boxtimes}x}$ and
${z'=l\stackrel{d}{\scriptstyle\boxtimes}z}$, then Eq.~\eqref{x'+z=x+z'} becomes ${l=k}$.
Thus, it validates the statement~\eqref{comm-power}.

Now we can build a set 
\begin{equation}
\label{commut-set}
\texttt{S}^{(x,z)}_d:=\big\{
X^{k\,\stackrel{d}{\scriptscriptstyle\boxtimes}\,x}
Z^{k\,\stackrel{d}{\scriptscriptstyle\boxtimes}\,z}\,:\,
k\in\mathbb{Z}_d\ \ \mbox{and}\ \ k\neq0\big\}
\end{equation}
of ${d-1}$ commuting operators (without any scalar multiple of $I$).
No operator of the Pauli basis~\eqref{Basis} outside of ${\texttt{S}^{(x,z)}_d\cup \{I\}}$ commutes with any operator of $\texttt{S}^{(x,z)}_d$ due to the result \eqref{comm-power}.
Note that an integral power of Pauli operator 
\begin{equation}
\label{(X^xZ^z)^k}
(X^xZ^z)^k=
\omega_d^{\frac{k(k-1)}{2}x\,\stackrel{d}{\scriptscriptstyle\boxtimes}\,z} X^{k\,\stackrel{d}{\scriptscriptstyle\boxtimes}\,x}
Z^{k\,\stackrel{d}{\scriptscriptstyle\boxtimes}\,z}
\end{equation}
commutes with all the elements of $\texttt{S}^{(x,z)}_d$ and
belongs to the group $\texttt{P}_d$, but not always to the basis~\eqref{Basis}. 
In particular,
\begin{equation}
\label{(X^xZ^z)^d}
(X^xZ^z)^d=
\begin{cases} 
I                 
& \text{if } d \text{ is odd prime}\\
(-1)^{x\,\stackrel{d}{\scriptscriptstyle\boxtimes}\,z}\, I       
& \text{if } d \text{ is even prime}
\end{cases}
\end{equation}
for every non-identity operator of the basis~\eqref{Basis}.
Due to the property~\eqref{(X^xZ^z)^d}, every Pauli operator generates a cyclic subgroup of $\texttt{P}_d$.
Besides, every cyclic group of order $d$ is isomorphic to $\mathbb{Z}_d$ \cite{Lidl86}.
With \eqref{(X^xZ^z)^d}, one can also realize
\begin{equation}
\label{commut-set-equ}
\texttt{S}^{(x,z)}_d=\texttt{S}^{\bm(l\,\stackrel{d}{\scriptscriptstyle\boxtimes}\,x\,,\,l\,\stackrel{d}{\scriptscriptstyle\boxtimes}\,z\bm)}_d\quad\mbox{for}\quad l=1,2,\cdots,d-1\,,
\end{equation}
and can compute eigenvalues of the Pauli operator
and then ${\text{tr}(X^xZ^z)=d\,\delta_{x,0}\delta_{z,0}}$ [see the orthogonality relation~\eqref{orth-XZ}].
Only a scalar multiple of the identity operator in $\texttt{P}_d$ have a nonzero trace, ${\text{tr}(I)=d}$, whereas all other operators are traceless.
By the way, eigenvalues of every non-identity ${X^xZ^z}$ are distinct powers of $\omega_d$, except for the case ${d=2}$ and ${x=1=z}$, where eigenvalues are ${\pm\,\text{i}}$.

Now we can split the Pauli basis~\eqref{Basis}, without the identity operator, into ${d+1}$ disjoint subsets
\begin{equation}
\label{d+1 subsets}
\Big\{\texttt{S}^{(1,0)}_d,\texttt{S}^{(1,1)}_d,
\texttt{S}^{(1,2)}_d,
\cdots,\texttt{S}^{(1,d-1)}_d\Big\}\cup\Big\{\texttt{S}^{(0,1)}_d\Big\}
\,,
\end{equation}
where each subset contains ${d-1}$ pairwise commuting operators \cite{Bandyopadhyay02}. As ${\texttt{S}^{(x,z)}_d}$ is constructed with ${X^xZ^z}$ [see \eqref{commut-set}], the sequence~\eqref{d+1 subsets} of subsets can be produced by the operators
\begin{equation}
	\label{d+1 XZ op}
	\big\{X,XZ,XZ^2,\cdots,XZ^{d-1}\big\}\cup \big\{Z\big\}\,.
\end{equation} 
Splitting~\eqref{d+1 subsets} of the basis~\eqref{Basis} is \emph{unique}, however $\texttt{S}^{(x,z)}_d$ can be stated differently [see \eqref{commut-set-equ}].

Since the unitary operators in $\texttt{S}^{(1,t)}_d$ (${t=0,\cdots,d-1}$) are---linearly independent but---functions of ${XZ^t}$ [see \eqref{(X^xZ^z)^k}] and all the eigenvalues of ${XZ^t}$ are distinct, one can find unique (up to a permutation of and global phases to the kets) orthonormal eigenbasis $\texttt{B}^{t}_d$ of ${XZ^t}$ for the whole set $\texttt{S}^{(1,t)}_d$.
In fact, our original basis~\eqref{B_i} is a joint eigenbasis of all the operators in $\texttt{S}^{(0,1)}_d$ that are integral powers of $Z$.
The orthonormal eigenbases
\begin{equation}
\label{d+1 bases}
\Big\{\texttt{B}^0_d\,,\,\texttt{B}^1_d\,,\,
\texttt{B}^2_d\,,\,\cdots\,,\,\texttt{B}^{d-1}_d\Big\}
\cup\Big\{\,\texttt{B}^{d}_d\,\Big\}\;,
\end{equation} 
associated with the disjoint subsets~\eqref{d+1 subsets} in the same order, compile a complete set of ${d+1}$ MUBs of $\mathscr{H}_{d}$ \cite{Ivanovic81,Bandyopadhyay02}.
These ${d+1}$ MUB-settings are sufficient for estimating any property---that is, the expectation value of any operator on $\mathscr{H}_{d}$---of a $d$-level quantum system.

Applying the Fourier operator $F$ [of Eq.~\eqref{F_i}] to $\texttt{B}^{d}_d$ we get $\texttt{B}^{0}_d$, and then applying \cite{Gottesman98,Gottesman99,Klimov05}
\begin{eqnarray}
\label{V}
V:=
\begin{cases} 
\textstyle\sum\nolimits_{j=0}^{d-1}\omega_{d}^{\,\frac{1}{2}j\,
	{\scriptscriptstyle\stackrel{d}{\scriptscriptstyle\boxtimes}}\,(j-1)}\,|j\rangle\langle j|                  
& \text{if } d \text{ is odd prime}\\
\textstyle\sum\nolimits_{j=0}^{d-1}\omega_{d}^{\,\frac{1}{2}j}\,|j\rangle\langle j|      
& \text{if } d \text{ is even prime}
\end{cases}\qquad\quad
\end{eqnarray}
${t\in\{0,\cdots,d-1\}}$ times to $\texttt{B}^{0}_d$ we obtain $\texttt{B}^{t}_d$.
$F$ and $V$ are also cyclic unitary operators:
\begin{equation}
\label{F V cycle}
I=
\begin{cases} 
F^4=V^d              
& \text{if } d \text{ is odd prime}\\
F^2=V^{2d}     
& \text{if } d \text{ is even prime}\,.
\end{cases}
\end{equation}
By observing that ${j(j-1)}$ represents even numbers, one can immediately see ${V^d=I}$ when $d$ is a odd prime number. Whereas one can directly inspect ${V^{d}=Z}$ for ${d=2}$.

Every unitary operator $U$ defines a linear transformation 
${A\stackrel{U}{\longrightarrow} UAU^\dagger}$, called unitary conjugation, on the space of operators $\mathscr{B}(\mathscr{H}_d)$.
With the Weyl commutation relation~\eqref{Weyl commutation}, one can 
realize
\begin{equation}
\label{Conjugation-relation-1}
\big(X^{x'}Z^{z'}\big)\,X^xZ^z\,\big(X^{x'}Z^{z'}\big)^\dagger
=\omega_d^{\,x\,\stackrel{d}{\scriptscriptstyle\boxtimes}\,z'-
	x'\,\stackrel{d}{\scriptscriptstyle\boxtimes}\,z}X^xZ^z\,,\qquad
\end{equation}
which reveals that a Pauli operator only introduces a phase factor to 
another Pauli operator.
By comparing Eq.~\eqref{Conjugation-relation-1} with Eqs.~\eqref{commute} and \eqref{x'+z=x+z'}, we can say:
if and only if the phase factor is 1 then the two operators commute and lie in the same set~\eqref{commut-set}.
When a Pauli operator is applied to one of the MUBs~\eqref{d+1 bases}, it either gives global phases to the kets
or permutes them or both \cite{Bandyopadhyay02}.
After a Pauli operation, the transformed basis essentially represents the same measurement setting (only the labels of outcomes get permuted).

With the conjugation relations~\eqref{FXF,FZF},
\begin{eqnarray}
\label{VXV,VZV}
VXV^\dagger=\begin{cases} 
XZ     & \text{if } d \text{ is odd}\\
\text{i}XZ     & \text{if } d \text{ is even}
\end{cases},\ \ \mbox{and}\ \ VZV^\dagger=Z\,,\ \quad\quad
\end{eqnarray}
we can recognize that $F$ and $V$ change one Pauli operator into other [see also Tables~\ref{tab:Conj, d=2} and \ref{tab:Conj, d=3}].
A group of such unitary operators---that map the Pauli group $\texttt{P}_d$ onto itself under the unitary conjugation---is so-called the Clifford group $\texttt{CF}_d$ \cite{Gottesman98,Gottesman99}, which of course contains $\texttt{P}_d$.
Moreover, $F$ and $V$ are non-Pauli Clifford operators that transform
one MUB of~\eqref{d+1 bases} into another (up to an order of and phase factors to the kets with in a basis).
For ${d=2,3}$, we can have all the members of $\texttt{CF}_d$ by multiplying $F$ and $V$ only \cite{Gottesman98,Gottesman99}.

\begin{table}[H]
	\centering
	\caption{For ${d=2}$, mapping of the Pauli operators under $F$ and $V$ conjugations, where ${Y=\text{i}XZ}$.}
	\label{tab:Conj, d=2}
	\begin{tabular}{lcr@{\hspace{2mm}} | @{\hspace{2mm}}lcr}
		\hline\hline\rule{0pt}{3ex}  
		&$F$&              &       &$V$& \\
		$X$&$\longrightarrow$&$Z$      &    $X$&$\longrightarrow$&$Y$  \\
		$Y$&$\longrightarrow$&$-Y$     &    $Y$&$\longrightarrow$&$-X$  \\
		$Z$&$\longrightarrow$&$X$      &    $Z$&$\longrightarrow$&$Z$  \\
		\hline\hline
	\end{tabular}
\end{table}

\begin{table}[H]
	\centering
	\caption{For ${d=3}$, mapping of the Pauli operators of the basis~\eqref{Basis} under $F$ and $V$ conjugations, where ${\omega=\exp(\text{i}\tfrac{2\pi}{3})}$.}
	\label{tab:Conj, d=3}
	\begin{tabular}{lcr@{\hspace{2mm}} | @{\hspace{2mm}}lcr}
		\hline\hline\rule{0pt}{3ex}  
		&$F$&              &       &$V$& \\
		$X$&$\longrightarrow$&$Z$      &    
		$X$&$\longrightarrow$&$XZ$     \\
		$X^2$&$\longrightarrow$&$Z^2$  &    
		$X^2$&$\longrightarrow$&$\omega X^2Z^2$  \\
		[1mm]
		$XZ$&$\longrightarrow$&$\omega^2X^2Z$  &    $XZ$&$\longrightarrow$&$XZ^2$  \\
		$X^2Z^2$&$\longrightarrow$&$\omega^2XZ^2$  & $X^2Z^2$&$\longrightarrow$&$\omega X^2Z$         \\ 
		[1mm]
		$XZ^2$&$\longrightarrow$&$\omega XZ$ &  
		$XZ^2$&$\longrightarrow$&$X$  \\
		$X^2Z$&$\longrightarrow$&$\omega X^2Z^2$ & 
		$X^2Z$&$\longrightarrow$&$\omega X^2$ \\ 
		[1mm]
		$Z$&$\longrightarrow$&$ X^2$      &    
		$Z$&$\longrightarrow$&$Z$  \\
		$Z^2$&$\longrightarrow$&$X$   & 
		$Z^2$&$\longrightarrow$&$Z^2$ \\
		\hline\hline
	\end{tabular}
\end{table}

Note that, for ${d=2}$, the Pauli group~\eqref{Pauli-gp} carries both Hermitian as well as skew-Hermitian operators, and \emph{no} unitary operator can map a nonzero Hermitian to a skew-Hermitian operator under the conjugation. Observe that we obtain ${\text{i}XZ=Y}$ in the conjugation~\eqref{VXV,VZV}.
Technically, $Y$ does not belong to the Pauli group defined by~\eqref{Pauli-gp}, but then one can build $\texttt{P}_2$ by multiplying $X,Y,$ and $Z$.
Therefore, for a qubit, we adopt the basis 
\begin{equation}
\label{d=2 P-basis}
\big\{\,I\,,\,X\,,\,Y\,,\,Z\,\big\}
\end{equation}
of $\mathscr{B}(\mathscr{H}_2)$ instead of \eqref{Basis}.

With the orthonormal basis~\eqref{B_i} of $\mathscr{H}_d$, we can build
another set of operators
\begin{equation}
\label{Basis-2}
\big\{|j\rangle\langle k|:j,k\in\mathbb{Z}_d\big\}\,,
\end{equation}
which also constitutes an orthogonal basis, like \eqref{Basis}, of the Hilbert-Schmidt space $\mathscr{B}(\mathscr{H}_d)$ since
\begin{equation}
\label{orth-|j><k|}
\big\lgroup\, |j'\rangle\langle k'|\,,\,|j\rangle\langle k|\,\big\rgroup_\textsc{hs}
=\langle j'|j\rangle\langle k|k'\rangle
=\delta_{j,j'}\,\delta_{k,k'}.
\end{equation} 
For ${j\neq k}$, ${|j\rangle\langle k|}$ is neither a unitary nor a Hermitian operator, and it is not even diagonalizable. 
Mathematically it is very useful because an operator is generally  expressed in terms of basis~\eqref{Basis-2} first [see $X$, $Z$, $F$, and $V$ defined above and examples in Sec.~\ref{sec:Examples}] and then in the Pauli basis~\eqref{Basis} through the relations
\begin{eqnarray}
\label{proj-op}
|j\rangle\langle k|
&=&X^j\,|0\rangle\langle 0|\,X^{-k}
=X^j\left[\tfrac{1}{d}
\textstyle\sum\nolimits_{z=0}^{d-1}Z^z\,
\right ] X^{-k}\nonumber\\
&=& \tfrac{1}{d}
\textstyle\sum\nolimits_{z=0}^{d-1}\, \omega_{d}^{-z\,\stackrel{d}{\scriptscriptstyle\boxtimes}\,k}\, X^{j-k}\,Z^z.
\end{eqnarray}
One can easily own these relations by exploiting \eqref{X_i}, \eqref{Z_i}, and \eqref{Weyl commutation}.


\subsection{For composite system}\label{app:composit Pauli Gp}

We begin here by recalling the prime factorization ${\mathsf{d}=\textstyle\prod\nolimits_{i=1}^{N}d_i}$
from Sec.~\ref{sec:criteria}. 
By putting $i$ in the subscript, we have mathematical objects for $i$th subsystem from the previous part.
With the Cartesian product of $\mathbb{Z}_{d_i}$ [taken from \eqref{Z_d}], we construct $N$-tuples  
\begin{equation}
\label{N-tuple}
\qquad
\mathsf{j}:=(\,j_1,\cdots, j_{\scriptscriptstyle N})
\;\in\;\mathcal{Z}_\mathsf{d}:=
\mathbb{Z}_{d_1}\times\cdots\times\mathbb{Z}_{d_N}\,,
\end{equation}
which are $\mathsf{d}$ in number.
Next we formulate the componentwise addition
\begin{equation}
\label{comp-add}
\mathsf{j}\,{\scriptstyle\boldsymbol{\boxplus}}\,\mathsf{j}':=
\Big(j_1\stackrel{d_1}{\scriptstyle\boxplus}j_1'\,,\,\cdots\,,\,
j_{\scriptscriptstyle N}\stackrel{d_{\scriptscriptstyle N}}{\scriptstyle\boxplus}j_{\scriptscriptstyle N}'\Big)\,.
\end{equation}

By merging orthonormal bases $\texttt{B}^{d_i}_{d_i}$ of $\mathscr{H}_{d_i}$ [defined as \eqref{B_i} for every subsystem $i$],
we have a \emph{product} orthonormal basis of $\mathscr{H}_\mathsf{d}$:
\begin{equation}
\label{pro-basis}
\mathcal{B}_0:=
\big\{|\,\textsf{j}\,\rangle:\textsf{j}\in\mathcal{Z}_\mathsf{d}\big\}\,,
\ \ \mbox{where}\ \ 
|\,\textsf{j}\,\rangle:=
\operatorname*{\otimes}_{i=1}^{N}|j_i\rangle,\ \
|j_i\rangle\in\texttt{B}^{d_i}_{d_i},
\end{equation}
and 
\begin{equation}
\label{orth-j}
\langle\,\textsf{j}'\,|\,\textsf{j}\,\rangle=
\delta_{\textsf{j},\textsf{j}'}=
\textstyle\prod\nolimits_{i=1}^{N}\delta_{j_i,j_i'}
\end{equation}
is then the orthonormality relation.
Basically, every $N$-tuple $\textsf{j}$ is created by the Cartesian product, and the associated state vector ${|\,\textsf{j}\,\rangle}$ by the tensor product.

With tensor product, we can also build product-Pauli operators
\begin{equation}
\label{X,Z N}
\textsf{X}_\textsf{x}:=\operatorname*{\otimes}_{i=1}^{N}X_i^{x_i}
\quad\mbox{and}\quad
\textsf{Z}_\textsf{z}:=\operatorname*{\otimes}_{i=1}^{N}Z_i^{z_i}
\end{equation}
for every ${\textsf{x},\textsf{z}\in\mathcal{Z}_\mathsf{d}}$. 
Local operators $X_i$ and $Z_i$ are---of period $d_i$---defined with the basis $\texttt{B}^{d_i}_{d_i}$
by Eqs.~\eqref{X_i} and \eqref{Z_i}. 
With these equations, action of $\textsf{X}_\textsf{x}$ and $\textsf{Z}_\textsf{z}$ on 
the product-basis $\mathcal{B}_0$ is described as
\begin{eqnarray}
\label{X,Z on B0}
\textsf{X}_\textsf{x}|\,\textsf{j}\,\rangle&=&|\,\mathsf{j}\,{\scriptstyle\boldsymbol{\boxplus}}\,\mathsf{x}\,\rangle\\
\textsf{Z}_\textsf{z}|\,\textsf{j}\,\rangle&=&
\Big(\textstyle\prod\nolimits_{i=1}^{N}\omega_{d_i}^{\,z_i\,\stackrel{d_i}{\scriptstyle\boxtimes}\,j_i}\Big)
|\,\textsf{j}\,\rangle
\end{eqnarray}
[see \eqref{comp-add} for the componentwise addition ${\mathsf{j}\,{\scriptstyle\boldsymbol{\boxplus}}\,\mathsf{x}}$].
By the Weyl commutation~\eqref{Weyl commutation}, we obtain
\begin{equation}
\label{Weyl commutation N}
\textsf{Z}_\textsf{z}\textsf{X}_\textsf{x}=
\Big(\textstyle\prod\nolimits_{i=1}^{N}\omega_{d_i}^{\,x_i\,\stackrel{d_i}{\scriptstyle\boxtimes}\,z_i}\Big)
\textsf{X}_\textsf{x}\textsf{Z}_\textsf{z}\,.
\end{equation}

Through the local Pauli basis~\eqref{Basis}, we have
the \emph{product}-Pauli basis 
\begin{equation}
\label{HS-basis N}
\big\{\Lambda^{(\textsf{x},\textsf{z})}:\textsf{x},\textsf{z}\in\mathcal{Z}_\mathsf{d}\big\}
\quad \mbox{with}\quad \Lambda^{(\textsf{x},\textsf{z})}=\textsf{X}_\textsf{x}
\textsf{Z}_\textsf{z}
\end{equation}
of $\mathsf{d}^2$-dimensional Hilbert-Schmidt space $\mathscr{B}(\mathscr{H}_\mathsf{d})$, where the orthogonality
relation
\begin{equation}
\label{orth-Lambda}
\big\lgroup\Lambda^{(\textsf{x}',\textsf{z}')}, \Lambda^{(\textsf{x},\textsf{z})}\big\rgroup_\textsc{hs}=
\textsf{d}\delta_{\textsf{x},\textsf{x}'}\delta_{\textsf{z},\textsf{z}'}=\textstyle\prod\nolimits_{i=1}^{N}
d_i\delta_{x_i,x_i'}\delta_{z_i,z_i'} 
\end{equation}
is drawn from \eqref{orth-XZ} as 
${\text{tr}(A\otimes B)=\text{tr}(A)\text{tr}(B)}$.
Actually, the above statement represents
\emph{Bohr's principle of complementarity}: ``For each quantum degree  of freedom (subsystem) there is a pair of complementary observables
($X_i,Z_i$)
and all observables are functions of this pair"---as quoted in \cite{Englert06}.
Furthermore, with Eqs.~\eqref{pro-basis} and \eqref{orth-j} as well as with
Eqs.~\eqref{HS-basis N} and \eqref{orth-Lambda}, one can appreciate the fact that tensor products of basis-elements associated with subsystems provide a basis for the composite system, which is also concluded by Wootters \cite{Wootters90} (see also \cite{Lawrence02}).

In general, a product-Pauli operator---does not have all distinct eigenvalues---is degenerate (especially, when a composite system carries two or more same level subsystems).
Consequently, a product operator can possess more than one eigenbasis, and some of these can be entangled bases [for example, see the Bell- and \textsc{ghz}-bases in Sec.~\ref{sec:Examples}].
If no operator in the tensor product~\eqref{pro-Pauli-op} is identity, then $\Lambda^{(\textsf{x},\textsf{z})}$ has a unique (up to a permutation of and global phase to the kets) orthonormal \emph{product} eigenbasis. For instance, $\mathcal{B}_0$ is a product-eigenbasis of
$\operatorname*{\otimes}_{i=1}^{N}Z_i$.
There are total 
\begin{equation}
\label{Tot-L-setts}
\textstyle\prod\nolimits_{i=1}^{N}(d_i+1)
\end{equation}
such product bases, each of them is specified by a set of local MUBs
\begin{eqnarray}
\label{pro basis}
\big\{\texttt{B}_{d_1}^{t_1}\,,\,\texttt{B}_{d_2}^{t_2}
\,,\,\cdots\,,\,\texttt{B}_{d_N}^{t_N}\big\}
&\equiv&\{t_1\,,\,t_2\,,\,\cdots\,,\,t_{\scriptscriptstyle N}\}\,,
\ \ \mbox{where}\qquad\quad\\ 
\label{ti}
t_i&\in&\{0,1,\cdots,d_i\}
\end{eqnarray}
is the index for MUBs, such as \eqref{d+1 bases}, attached to $i$th subsystem.
Here, a \emph{single-setting for local-measurements} is completely laid out by the collection~\eqref{pro basis} of MUB-indices.
Moreover, tensor products of the kets---one ket at a time taken from each basis of the setting~\eqref{pro basis}---compose 
the associated product basis, like $\mathcal{B}_0$ defined in \eqref{Basis-2 N}.
Since a pair of local settings can share one or more local MUBs, only  certain, not all, pairs of the product bases are mutually unbiased.

Two product-Pauli operator $\Lambda^{(\textsf{x}',\textsf{z}')}$ and $\Lambda^{(\textsf{x},\textsf{z})}$ 
can commute even if they do not commute componentwise, that is  (locally)
\begin{equation}
\label{x'+z=x+z' N}
x_i\stackrel{d_i}{\scriptstyle\boxtimes}z_i'=x_i'\stackrel{d_i}{\scriptstyle\boxtimes}z_i
\quad(\text{mod}\ d_i)\quad\mbox{for every}\quad i=1,\cdots,N\,.
\end{equation}
All the Pauli operators appear in the decomposition of the Bell-projector~\eqref{B-proj-Pauli} commute with each other, but not all of them commute componentwise [see also the \textsc{ghz}-projector~\eqref{|ghz><ghz|}]. 
\begin{equation}
\label{comm-power N}
\parbox{0.8\columnwidth}
{Expectation values $\langle\Lambda^{(\textsf{x}',\textsf{z}')}\rangle_\rho$ and $\langle\Lambda^{(\textsf{x},\textsf{z})}\rangle_\rho$ of the two product operators can be estimated with a \emph{single local}-MUB setting---such as~\eqref{pro basis}---if and only if the two operators commute \emph{componentwise} (see Observation~1 in \cite{Toth05b}), that is they follow \eqref{x'+z=x+z' N}, otherwise we need two separate settings.  
}
\end{equation}
We adopt this rule to count the number of local settings needed to estimate the expectation value of an entangled projector in Sec.~\ref{sec:Examples}.

For $i$th subsystem, a Clifford operator is a composition of operators such as $F_i$ and $V_i$ [defined by Eqs.~\eqref{F_i} and \eqref{V} with the basis $\texttt{B}_{d_i}^{d_i}$], which are adequate to deliver all
local Clifford operators as long as ${d_i=2,3}$.
When ${d_i>3}$, we might need one more operator---denoted by $S_a$ and described by the mappings (21) and (22) in \cite{Gottesman99}---to generate the complete (local) Clifford group $\texttt{CF}_{d_i}$.
The product-Clifford group $\mathcal{C}_\mathsf{d}$ is produced by the tensor products of local Clifford operators, just like its subgroup $\mathcal{P}_\mathsf{d}$.

We are only considering the elements of $\mathcal{C}_\mathsf{d}$ for the unitary conjugation in order to gain more conditions for the entanglement detection.
Note that, for a composite number $\mathsf{d}$, $\mathcal{C}_\mathsf{d}$ is not the complete Clifford group, which also contains non-product operators such as \eqref{bell-U}, \eqref{ghz-U}, and \eqref{h-U}.
One product-Pauli operator transforms other as
\begin{equation}
\label{Conjugation-relation-N}
\Lambda^{(\textsf{x}',\textsf{z}')}\Lambda^{(\textsf{x},\textsf{z})}
{\Lambda^{(\textsf{x}',\textsf{z}')}}^\dagger
=
\Big(\textstyle\prod\nolimits_{i=1}^{N}
\omega_{d_i}^{\,x_i\,\stackrel{d_i}{\boxtimes}\,z_i'-x_i'\,\stackrel{d_i}{\boxtimes}\,z_i}\Big)
\Lambda^{(\textsf{x},\textsf{z})}\,,
\end{equation}
which is derived from the (local) conjugation relation~\eqref{Conjugation-relation-1}.
For this paper, Eqs.~\eqref{FXF,FZF}, \eqref{VXV,VZV} and Tables~\ref{tab:Conj, d=2}, \ref{tab:Conj, d=3} are sufficient to have all the Clifford conjugations (see \cite{Gottesman98,Gottesman99} for further details).

For the sake of completeness, we supply the orthonormal operator basis
\begin{equation}
\label{Basis-2 N}
\big\{\,|\textsf{j}\rangle\langle \textsf{k}|:
\textsf{j},\textsf{k}\in\mathcal{Z}_\textsf{d}\,\big\}\;,
\end{equation} 
whose members are related to the elements of product-Pauli basis~\eqref{HS-basis N} as  
\begin{eqnarray}
\label{proj-op N}
|\textsf{j}\rangle\langle \textsf{k}|&=&
\operatorname*{\otimes}_{i=1}^{N}|j_i\rangle\langle k_i|\nonumber\\
&=&
\operatorname*{\otimes}_{i=1}^{N}\left[\tfrac{1}{d_i}
\textstyle\sum\nolimits_{z_i=0}^{d_i-1}\, \omega_{d_i}^{-z_i\,\stackrel{d_i}{\boxtimes}\,k_i}\, X_i^{j_i-k_i}\,Z_i^{z_i}
\right]\nonumber\\
&=&\tfrac{1}{\mathsf{d}}
\textstyle\sum\limits_{\textsf{z}\,\in\,\mathcal{Z}_\textsf{d}}
\Big(\textstyle\prod\nolimits_{i=1}^{N}
\omega_{d_i}^{-z_i\,\stackrel{d_i}{\boxtimes}\,k_i}\Big)
\Lambda^{(\textsf{j}-\textsf{k},\textsf{z})}\,.
\end{eqnarray}
Like the addition~\eqref{comp-add}, ${\textsf{j}-\textsf{k}}$ is the componentwise subtraction, and the above transformation is acquired from \eqref{proj-op}.

${}$



\begin{thebibliography}{99}
\bibitem{note}
		Without loss of generality, we choose the same orthonormal basis~\eqref{B_i} of $\mathscr{H}_d$ for every qudit while defining the entangled kets in Sec.~\ref{sec:Examples}. 
		One can very well select distinct bases for individual qudits---to begin with a completely general product basis~\eqref{pro-basis}---then obviously the $X$ operator for one qudit will not be the same for other, and that goes for any local operator.
		However, it does not affect the total number of conditions and their forms for the entanglement detection.
\end{thebibliography}
\end{document}